\documentclass[a4paper,11pt]{article}
\pdfoutput=1 

\usepackage{jheppub} 

\usepackage[T1]{fontenc} 

\newcommand{\alp}{\alpha^\prime}

\newcommand{\Amp}{\mathcal{A}}
\newcommand{\de}{\partial}
\newcommand{\produtt}{\prod_{i<j}z_{ij}^{2\alp p_ip_j}}

\newcommand{\tr}{\text{tr}}
\newcommand{\lef}{\langle}
\newcommand{\re}{\rangle}
\newcommand{\al}{\alpha}
\newcommand{\ald}{\dot{\alpha}}
\newcommand{\betad}{\dot{\beta}}

\newcommand{\ap}{\alpha^\prime}
\newcommand{\ie}{{\it{i.e.~} }}
\newcommand{\eg}{{\it e.g.~}}
\newcommand{\be}{\begin{equation}}
\newcommand{\ee}{\end{equation}}
\newcommand{\bea}{\begin{equationarray}}
\newcommand{\eea}{\end{equationarray}}
\newcommand{\bit}{\begin{itemize}}
\newcommand{\eit}{\end{itemize}}

\newcommand{\cA}{\mathcal{A}}
\newcommand{\cB}{\mathcal{B}}

\newcommand{\cH}{\mathcal{H}}

\newcommand{\cM}{\mathcal{M}}
\newcommand{\cN}{\mathcal{N}}
\newcommand{\cO}{\mathcal{O}}
\newcommand{\cP}{\mathcal{P}}
\newcommand{\cQ}{\mathcal{Q}}

\newcommand{\cS}{\mathcal{S}}

\title{\boldmath On the soft limit of open string disk amplitudes \\ with massive states}

 \author{M.~Bianchi}
 \author{and A.~L.~Guerrieri}
 \affiliation{Dipartimento di Fisica, Università di Roma "Tor Vergata" \\ and Sezione INFN  di Roma II "Tor Vergata" \\ Via della Ricerca Scientifica, 00133 Rome, Italy}

\emailAdd{massimo.bianchi@roma2.infn.it}
\emailAdd{andrea.guerrieri@roma2.infn.it}

\abstract{We discuss the soft behaviour of open string amplitudes with gluons and massive states in any dimension.  Notwithstanding non-minimal couplings of massive higher spin states to gluons, relying on OPE and factorization, we argue that the leading and sub-leading terms are universal and identical to the ones in Yang-Mills theories. In order to illustrate this, we compute some 4-point amplitudes on the disk involving gluons, massive states and, for the bosonic string, tachyons.  For the superstring, we revisit the structure of the massive super-multiplets at the first massive level and rewrite the amplitudes in $D=4$ in the spinor helicity formalism, that we adapt to accommodate massive higher spin states. We also check the validity of a recently obtained formula relating open superstring amplitudes for mass-less states at tree-level to SYM amplitudes, by factorisation on two-particle massive poles. Finally we analyse the holomorphic soft limit of superstring amplitudes with one massive insertion.}

\begin{document} 
\maketitle
\flushbottom
\section{Introduction and motivations}
Recently the soft behaviour of scattering amplitudes has received renewed attention in connection with the extended BvBMS symmetry \cite{Bondi:1962px, Barnich:2009se, Barnich:2011ct, Barnich:2011mi, Strominger:2013jfa, He:2014laa}.
It has been long  known that gauge theory and gravity amplitudes expose universal behaviours when one of the mass-less external momenta is `soft' \ie $k \rightarrow 0$ \cite{Low:1954kd, Weinberg:1965nx, Gross:1968in}. In both cases the leading behaviour is singular, \ie goes as $\delta^{-1}$ if $k=\delta \hat{k}$ with $\hat{k}$ some fixed momentum \cite{Low:1954kd}. The sub-leading terms can be derived from the leading ones and are largely fixed by gauge invariance. In particular, in gauge theories the sub-leading behaviour $\delta^{0}$ is universal, too. In gravity not only the sub-leading behaviour $\delta^{0}$ but also the next-to-subleading or sub-sub-leading behaviour $\delta^{+1}$ is universal~\cite{Weinberg:1965nx,Gross:1968in, Cachazo:2014fwa}. 

The problem of what happens when loops or non-minimal higher derivative couplings are included was addressed in \cite{Bern:2014oka, He:2014bga, Cachazo:2014dia, Bianchi:2014gla, Larkoski:2014bxa}.  At the loop level IR divergences tend to spoil the analysis. Yet, in supersymmetric theories such as $\cN=4$ SYM, one can define loop integrands recursively and check that they expose the expected soft behaviour at all loops and for any choice of (massless) external legs. This may be viewed as a further constraint on (loop) amplitudes derived without resorting to standard perturbative methods 
(see \eg  \cite{Elvang:2013cua}  for a recent pedagogical review). 

When non-minimal interactions are considered, the result depends on the specific choice \cite{Bianchi:2014gla}. In gauge theories, $F^3$ terms do not change the universal soft behaviour of minimal coupling, while  $\phi F^2$ do modify even the leading term when $\phi$ is a massless scalar. Similarly, in gravity theories $R^3$ terms do not change the universal soft behaviour of minimal coupling, while  $\phi R^2$ do modify even the leading term when $\phi$ is a massless scalar such as the dilaton. 

These results are largely independent of the number of space-time dimensions and in particular apply to string theory in critical dimension and in lower dimensions \cite{Casali:2014xpa, Schwab:2014xua, Afkhami-Jeddi:2014fia, Adamo:2014yya, Schwab:2014fia, Bern:2014vva}. One has to distinguish between open and closed strings and between bosonic, super and heterotic strings. In \cite{Bianchi:2014gla}, the soft behaviour has been shown to be governed by the OPE of the vertex operators. As a result both open and closed superstring amplitudes with external massless states expose the expected soft behaviour, while closed bosonic string amplitudes don't, due to the tree-level non minimal coupling $\phi R^2$ with the dilaton. Open bosonic string amplitudes behave universally despite the presence already at tree level of the non-minimal $F^3$ correction to the standard Yang-Mills coupling and the coupling $T F^2$ to  the tachyon\footnote{Couplings to gravitons and other closed string states appear at 
higher order in the string coupling $g_s$.}. For the heterotic string at tree level, the soft behaviour of massless vector bosons is universal, since the trilinear coupling is purely of Yang-Mills type, while the soft behaviour of the graviton is non-universal due to $\phi R^2$ coupling\footnote{M.~B. would like I.~Antoniadis to stressing the tree level origin of this term in the heterotic string, which only is generated at one-loop in 4-dim Type II theories with 16 supercharges, such as after compactification on $K3\times T^2$}. No $R^3$ term however appears due to supersymmetry.

Aim of the present paper is to extend the analysis to open string amplitudes with massive external states 
(in the bosonic string case we will also consider tachyons as external states). 
Amplitudes with massive external states have not received much attention in the literature. See however \cite{Bianchi:2010dy, Bianchi:2010es, Bianchi:2011se, Black:2011ep, Skliros:2009cs, Skliros:2011si} for recent work on the subject and \cite{Dudas:1999gz, Chialva:2005gt, Bianchi:2006nf, Feng:2010yx, Feng:2011qc} and the review \cite{Anchordoqui:2009ja} for mass-less amplitudes with emphasys on the exchange of massive higher spin states.
 
The analysis is interesting in two respects. On the one hand couplings of string states are generically non-minimal, although probably unique. On the other hand, gravity and gauge interactions emerge quite naturally in string theory and one would expect the soft behaviour of scattering amplitudes to expose some universality thanks to gauge invariance. 

Plan of the paper is as follows. 
We start with open bosonic strings. After reviewing tri-linear couplings on the disk of tachyons, vector bosons and higher spin massive states, we compute some explicit 4-point amplitudes involving tachyons and massive states.  
We then consider open superstrings on the disk and perform a similar analysis in an arbitrary number of dimension $D\le 10$. For convenience and for comparison with the existing literature we rewrite superstring amplitudes in $D=4$ in the spinor helicity formalism, that we adapt to accommodate massive higher spin states, after revisiting the structure of the massive super-multiplets. 
We also check the validity of a recently obtained formula relating open superstring amplitudes for mass-less states to SYM amplitudes at tree-level \cite{Mafra:2011nv, Mafra:2011nw}, by factorizing 5-point amplitudes on the first massive pole and recovering our previous formulae. We explain how to generalise this procedure to an arbitrary number of massive 
external states. We then discuss the soft behaviour of open string amplitudes with gluons and massive states in any dimension and argue that the leading and sub-leading terms are universal and identical to the SYM case, relying on OPE and factorization. We then check this explicitly for the amplitudes, we previously computed.
We also analyse the holomorphic soft limit of superstring amplitudes with one massive state and check consistency with our expectations.  
Finally we will draw our conclusions and identify interesting directions for future investigation.  
Various appendices contain technical details that are included for completeness. 

\section{Open bosonic string amplitudes}

In order to check the soft behavior of four-point amplitudes on the disk in the open bosonic string, 
we summarize in Appendix~\ref{subsec:3bosonic} all the possible tri-linear couplings involving the vertex operators up to the first massive level.

\subsection{Vertex operators}

Our first goal is to compute scattering amplitudes with the insertion of vertex operators up to the first massive level. 
Up to normalization factors, the tachyon vertex operator is
\begin{align}
V_{T}=e^{ipX} \quad p^2=-m^2_{T}=\frac{1}{\alp};
\label{eq:tac}
\end{align}
the gluon vertex operator is
\begin{equation}
V_A=a_{\mu}\,i\partial X^{\mu}\, e^{ikX} \quad k^2=-m^2_A=0 \quad a{\cdot} k=0
\label{eq:vec}
\end{equation}
and the first massive level vertex operator is
\begin{align}
V_{H}&=H_{\mu\nu}\,i\de X^\mu\,i\de X^\nu\,e^{ipX}\quad p^2=-m^2_{H}=-\frac{1}{\alp}\nonumber\\
&H_{\mu\nu}=H_{\nu\mu}\quad H_{\mu\nu}p^{\mu}=0\quad \eta^{\mu\nu}H_{\mu\nu}=0.
\label{eq:H}
\end{align}
While the choice of the tachyon vertex operator is essentially unique, the choice of the vertex operators for the massless gluon $A$ 
and for the massive state ${H}$ is not unique. It is always possible to add the null operator $c\, ik\partial X e^{ikX} = \delta_{_{BRST}} e^{ikX}$ to $c\,V_A$. 
For ${H}$ one can choose a linear combination of the operator $V_B=B_\mu i\de^2 X^\mu e^{ipX}(z)$ and 
a generic $\widetilde{{H}}(z)=\widetilde{H}_{\mu\nu}i\de X^\mu i\de X^\nu e^{ipX}(z)$, with $\widetilde{H}$ an arbitrary two-index symmetric tensor. 
Nonetheless, due to BRST invariance, one has the freedom to fix the gauge in which $B_\mu=0$ and $H_{\mu\nu}$ is symmetric, traceless and traverse as in Eq.~(\ref{eq:H}).

\subsection{Chan-Paton factors and twist symmetry}
\label{sec:ChanPaton}

Although we will mostly consider `color-ordered' amplitudes on the disk, we would like to review some relevant aspect of the group theory structure.  
Disk amplitudes are cyclically invariant and can be dressed with Chan-Paton factors \cite{Paton:1969je}
\be
\cA(1,2,\ldots n) \rightarrow \widehat\cA(1,2,\ldots n) = \cA(1,2,\ldots n) \tr(t^1\ldots t^n)
\ee
where $t_a$ with $a=1, \ldots N^2$ are the generators of $U(N)$\footnote{We will not consider unoriented projections or symmetry breaking that may produce matter in bi-fundamental or (anti-)symmetric representations of the gauge group.}.
In modern terms this corresponds to the fact that open strings carry multiplicities associated to the D-branes they end on. $\cA(1,2,\ldots n)$ are called `color-stripped' or `color-ordered' amplitudes or simply sub-amplitudes. They enjoy such remarkable properties as \cite{Mangano:1990by}
\bit
\item Cyclic symmetry: $\cA(k,k{+}1,\ldots,n,1,2,\ldots k{-}1) = \cA(1,2,\ldots n)$
\item Twist symmetry: $\cA(n,n{-}1,\ldots, 2,1) = \prod_{i=1}^n \omega_i \cA(1,2,\ldots,n{-}1, n)$
\item Dual identity: $\cA(1,2,\ldots n) + \cA(2,1,3,\ldots n) + \ldots + \cA(2,3,\ldots,n{-}1,1,n) = 0$
\eit
where $\omega_{{S}} = {\pm}1$ is the eigenvalue in the state ${S}$ of the `twist' operator $\Omega$, that exchanges the two ends of the strings. In particular $\omega_{A} = -1$ while $\omega_{{T},{H}} = +1$. In general 
$\omega_{{S}} = (-1)^{N_{S}}$ where $N_{S}$ is the level of ${S}$. Pretty much as in gauge theory, complete amplitudes are obtained by summing over non-cyclical permutations of color-dressed amplitudes.
At 3-points one simply has  
\be 
\widehat\cA(1,2,3) = \cA(1,2,3)\tr(t^1t^2t^3) + \cA(1,3,2)\tr(t^1t^3t^2).
\ee
For the amplitude with three tachyons $\cA(1,3,2) = \cA(3,2,1) = + \cA(1,2,3)$, so that 
\be
\widehat\cA(1,2,3) = \cA(1,2,3)\tr(t^1t^2t^3 +t^1t^3t^2 ) = \cA(1,2,3)d^{123}, 
\ee
while for three vectors $\cA(1,3,2) = \cA(3,2,1) = - \cA(1,2,3)$, so that 
\be
\widehat\cA(1,2,3) = \cA(1,2,3)Tr(t^1t^2t^3 -t^1t^3t^2 ) = \cA(1,2,3)f^{123}. 
\ee
In general one gets $d^{abc}$ (`anomaly coefficients' or cubic Casimir) when 
$\prod_{i=1}^3 \omega_i = +1$ and $f^{abc}$ (structure constants) when 
$\prod_{i=1}^3 \omega_i = -1$. In particular all couplings $\widehat\cA(S,S,A)\sim f^{abc}$ whichever the state ${\cS}$. Moreover, at least for totally symmetric tensors in the first Regge trajectory, the dominant term at low energy is 
\be 
\cA(S_1,S_2,A_3) = f_{123} S_1^{\mu_1\ldots \mu_s} S_{2,\mu_1\ldots \mu_s} a_3{{\cdot}}(p_1-p_2) + \ldots
\ee
\emph{i.e.} string theory tries to be as `minimal' as it can! Yet there are higher derivative corrections to this, as we will see momentarily.

\subsection{Four-point bosonic string amplitude}

In this section we collect some open bosonic string amplitudes involving massless, massive and tachyonic states. Details of the computations can be found in Appendix B4. For simplicity we consider color-ordered amplitudes. Complete amplitudes arise after multiplying by the relevant Chan-Paton factors $\tr(t^1t^2t^3t^4)$ and summing over non-cyclic permutations. In fact, exploiting `twist symmetry' \ie $\Omega$ invariance, one can further reduce the sum to three terms (instead of six). For notational simplicity we will drop all adimensional constants (including powers of $g_s$) that are irrelevant for our analysis and the $\delta$-function of momentum conservation $(2\pi)^D \delta^D(\sum_i p_i)$ will be understood. 
To help recognising the light-like momenta we will denote them by $k$'s, while tachyonic and massive momenta will be denoted by $p$'s.
Starting with Veneziano amplitude (four tachyons)
\begin{equation}
\Amp({T}_1,{T}_2,{T}_3,{T}_4)=\frac{\Gamma(-1-\alp s)\Gamma(-1-\alp t)}{\Gamma(-2-\alp (s+t))}
\end{equation}
where $s=-(p_1+p_2)^2=-(p_3+p_4)^2$, $t=-(p_2+p_3)^2=-(p_1+p_4)^2$, $u=-(p_1+p_3)^2=-(p_2+p_4)^2$, with $s+t+u = - 4/\ap$, 
it is not difficult to compute the three-tachyons one-vector amplitude
\be
\Amp(A_1,{T}_2,{T}_3,{T}_4) =\frac{1}{\sqrt{2\alp}}\left(\frac{a_1p_2}{ k_1p_2} - \frac{a_1p_4}{ k_1p_4} \right)\frac{\Gamma(1+2\alp k_1p_2)\Gamma(1+2\alp k_1p_4)}{\Gamma(1-2\alp k_1p_3)} ,
\ee
the two-tachyons two-vectors amplitude
\begin{align}
&\Amp(A_1,A_2,{T}_3,{T}_4) = (\widetilde a_1\widetilde a_2+2\alp\widetilde a_1p_3\,\widetilde a_2p_3) \,\frac{\Gamma(1+2\alp k_1p_4)\Gamma(-1+2\alp k_1k_2)}{\Gamma(-2\alp k_1p_3)},
\end{align}
where 
\be
\widetilde a_i=a_i-\frac{a_ip_4}{k_ip_4}k_i,\quad i=1,2,
\ee
satisfy $\widetilde a_ip_4=0$,
and finally the two-tachyons-vector-tensor amplitude
\begin{align}
&\Amp({T}_1,{T}_2,A_3,{H}_4) = \frac{1}{\sqrt{2\alp}} \frac{\Gamma(-1+2\alp k_3p_4)\Gamma(1+2\alp p_2k_3)}{\Gamma(-2\alp p_1k_3)} \biggr[-2a_3Hp_2-2a_3Hk_3\frac{1+2\alp k_3p_1}{2-2\alp k_3p_4} \nonumber\\&+2\alp a_3p_4\biggr(p_2Hp_2\frac{1-2\alp k_3p_4}{2\alp k_3p_1}  +k_3Hk_3\frac{1+2\alp k_3p_1}{2-2\alp k_3p_4}+2p_2Hp_3\biggr) \nonumber\\
&-2\alp a_3p_2\biggr(\frac{k_3p_4\,\,p_2Hp_2}{2\alp p_2k_3\,\,p_1k_3}(1-2\alp k_3p_4)
-p_3Hp_3\frac{1+2\alp p_1k_3}{2\alp p_2k_3}-2p_2Hk_3\frac{1-2\alp k_3p_4}{2\alp p_2k_3}\biggr)\biggr] \: .\nonumber
\label{eq:TTVH}
\end{align}
Later on we will check that they enjoy the expected behavior in the soft limit. 

\section{Open superstring amplitudes}

\subsection{Vertex operators}

In this section we consider open superstring amplitudes involving gluons and massive states. At the first massive level, two independent 
string excitations appear: a symmetric, transverse and traceless tensor ${H}$ with $d_{{H}}=D(D-1)/2 -1$ degrees of freedom ($d_{{H}} = 44$ in $D=10$) and a totally antisymmetric transverse tensor ${C}$ with $d_{{C}} = (D-1)(D-2)(D-3)/6$ degrees of freedom ($d_{{H}} = 84$ in $D=10$).
It is worth to notice that in $D=4$, the tensor ${H}$ corresponds to a massive spin 2 particle, while the ${C}$ corresponds to a massive pseudo-scalar.
Up to normalization factors, In the canonical $q=-1$ super-ghost picture their vertex operators are
\begin{align}
V_A^{(-1)}&=e^{-\varphi}\,a{\cdot} \psi \,e^{ikX}\qquad k^2=0 \qquad k{\cdot} a=0\\
V_{H}^{(-1)}&=H_{\mu\nu}\,e^{-\varphi}\,i\de X^{\mu}\,\psi^{\nu}\,e^{ipX}\qquad \ap p^2 = {-}1 \qquad p_\mu H^{\mu\nu}=0\qquad H^\mu_\mu=0\\
V_{C}^{(-1)}&=C_{\mu\nu\rho}\,\psi^\mu\psi^\nu\psi^\rho \,e^{ipX}\qquad \ap p^2 = {-}1 \qquad  p_\mu C^{\mu\nu\rho}=0 \: .
\end{align}

For our purposes it is necessary to consider also vertex operators in the $q=0$ super-ghost picture 
\begin{align}
V_A^{(0)}&=\left(a{\cdot}i\de X + k{\cdot}\psi \,\, a{\cdot}\psi\right)e^{ikX};\\
V_{H}^{(0)}&=H_{\mu\nu}\,\left[ i\de X^\mu \left(i\de X^\nu+ p\cdot\psi \psi^\nu\right] + \de \psi \psi \right)\, e^{ipX};\\ 
V_{C}^{(0)}&=C_{\mu\nu\rho}\,e^{-\varphi} \left[i\de X^\mu + p{\cdot}\psi\psi^\mu\right]\psi^\rho\psi^\sigma \,e^{ipX}.
\end{align}

Higher spin massive states in the first Regge trajectory are described by vertex operators of the form 
 \be
V_{H_s}=H_{\mu_1 \dots \mu_s} \left[ \prod_{i=1}^s i\de X^{\mu_i} + p\psi\,\psi^{\mu_1}\prod_{i=2}^{s} i\de X^{\mu_i} + (s-1)\de \psi^{\mu_1}\psi^{\mu_2}\prod_{i=3}^{s}i\de X^{\mu_i} \right] \,e^{ipX}
 \ee 
with $\ap p^2 = (1{-}s)$ and $H$ totally symmetric, transverse and trace-less.
Their tri-linear couplings to the vector bosons schematically read
\begin{align}
\Amp(A_1,A_2,H_s)&=\langle c\,e^{-\varphi} V_A(z_1) \,c\,e^{-\varphi} V_A(z_2)\,c\,V_{H_s}(z_3)\rangle = (2\ap)^{s/2}(f_1f_2)^{\mu_1\mu_2}H_{\mu_1\dots\mu_s}\prod_{i=3}^s {k_{12}^{\mu_i}}
  \end{align}
  
In $D=4$ the above expressions drastically simplify if one resorts to the spinor helicity formalism and adapt it so as to encompass massive states.

\subsection{Supersymmetry}

Although we will only consider bosonic states in the NS sector of the open superstring, it is worth discussing the structure of the super-multiplet at the first massive level \cite{Koh:1987hm}.

In addition to the NS states ${H}$ and ${C}$ we have two spin 3/2 fermions of opposite chirality (in $D=10$) that combine to give a massive spin 3/2 fermion. In the canonical $q=-1/2$ super-ghost picture their vertex operators read
\be
V_{\Psi} = \Psi_\mu^\alpha S_\alpha \de X^\mu e^{-\varphi/2} e^{ipX}
\ee
and  
\be
V_{\widetilde\Psi} =\widetilde\Psi^\mu_\alpha W^\alpha_\mu e^{-\varphi/2} e^{ipX}
\ee
where $S_\alpha$ is a spin field of conformal dimension 5/8 in the ${\bf 16}$ irrep of $SO(1,9)$ and $W^\alpha_\mu = {:}C^\alpha \psi^\mu{:}$ is an excited spin field of conformal dimension 13/8 in the ${\bf 144'}$ irrep of $SO(1,9)$. BRST invariance implies transversality 
$p^\mu \Psi_\mu^\alpha = 0 = p_\mu \widetilde\Psi^\mu_\alpha$, $\Gamma$-traceleness 
$\Gamma^\mu_{\alpha\beta} \Psi_\mu^\beta = 0 = \Gamma_\mu^{\alpha\beta} \widetilde\Psi^\mu_\beta$ as well as 
\be
\Gamma^\mu_{\alpha\beta} p_\mu \Psi_\nu^\beta = i M \widetilde\Psi_{\nu,\alpha}
\quad , \quad
\Gamma_\mu^{\alpha\beta} p^\mu \widetilde\Psi_{\nu,\beta} = i M \Psi_\nu^\alpha  
\ee

The $\cN = (1,0)$ supersymmetry charge in $D=10$  is the gaugino vertex at zero momentum
\be
\cQ^{(-1/2)}_\alpha = \int dz S_\alpha e^{-\varphi/2}   
\ee 
In the $q=+1/2$ super-ghost picture one has 
\be
\cQ^{(+1/2)}_\alpha = \int dz \Gamma^\mu_{\alpha\beta} C^\beta \de X_\mu e^{+\varphi/2}   
\ee 

Acting with $\cQ^{(-1/2)}_\alpha$ on $V_{\Psi}$ and $V_{\widetilde\Psi}$ one gets combinations of the NS vertex operators $V_{C}$ and $V_{H}$ as well as the `auxiliary' vertices $V_B$ and $V_E$ in the canonical $q=-1$ picture, yielding very schematically\footnote{See \cite{Koh:1987hm} for the precise coefficients.} 
\be 
\delta H_{\mu\nu} = \varepsilon\Gamma_{(\mu} \Psi_{\nu)}
\qquad \delta C_{\mu\nu\rho} = \varepsilon \Gamma_{[\mu\nu} \widetilde\Psi_{\rho]}
\qquad 
\delta B_{\mu\nu} = \varepsilon\Gamma_{[\mu} \Psi_{\nu]}
\qquad
\delta E_{\mu} = \varepsilon  \widetilde\Psi_{\mu}
\ee

Similarly acting with $\cQ^{(+1/2)}_\alpha$ on the NS vertex operators $V_{C}$ and $V_{H}$ as well as on the `auxiliary' vertices $V_B$ and $V_E$ yields very schematically\footnote{See \cite{Koh:1987hm} for the precise coefficients.}
\be
\delta\Psi_\mu = \varepsilon\Gamma_{\lambda} \left[p^\lambda (H_{\mu\nu} + B_{\mu\nu} ) \Gamma^{\nu} + M \Gamma_\mu E^\lambda \right]+ 
M C_{\mu\nu\rho} \Gamma^{\nu\rho} \varepsilon + \ldots
\ee
and
\be
\delta\widetilde\Psi_{\mu} = \varepsilon\Gamma^\nu \left[M (H_{\nu\mu} + B_{\nu\mu} ) + p_\nu E_\mu \right] + \varepsilon \Gamma^\nu p_\nu C_{\mu\lambda\rho} \Gamma^{\lambda\rho} + \ldots \ee

\subsection{Dimensional reduction to $D=4$}

For obvious reasons we are particularly interested in the dimensional reduction to $D=4$.
The massive $\cN = (1,0)$ super-multiplet in $D=10$ at the first level yields a long multiplet of the $\cN = 4$ super-algebra  
\be
\{H_{\mu\nu}, 8\, \psi_\mu, 27\, Z_\mu, 48 \,\chi, 42 \,\varphi\}
\ee
comprising 128 bosonic and as many fermionic states. In order not to burden the notation $\mu, \nu, \ldots$ are now 4-dim indices, while $i,j, \ldots$ denote the internal 6 dimensions. The origin of the bosonic fields is as follows
\be
H_{\mu\nu} \leftarrow H_{\mu\nu}
\ee
\be
27 \,Z_\mu \leftarrow 6 \,H_{\mu, i}, 15 \,C_{\mu, ij}, 6 \,C_{\mu\nu,i}
\ee
since a massive vector in $D=4$ is equivalent to a massive anti-symmetric tensor,
and
\be
42\, \varphi \leftarrow 21 \,H_{ij}, 20 \,C_{ijk},  C_{\mu\nu\rho}
\ee
It is perhaps not surprising that these be in one-to-one correspondence with the (bosonic) fields in the 
$\cN = 4$ super-current multiplet, upon dualizing the six massive $H_{\mu, i}$ into as many massive anti-symmetric tensors $\widetilde{H}_{\mu\nu,i} = \varepsilon_{\mu\nu\lambda\rho} p^\lambda H^\rho_{i}/M$.
It is amusing to decompose this massive multiplet into massive multiplets of the $\cN = 1$ super-algebra 
\begin{align}
&\{H_{\mu\nu}, 8\, \psi_\mu, 27 \,Z_\mu, 48\, \chi, 42 \,\varphi\} \rightarrow \nonumber\\
&\{H_{\mu\nu}, 2\, \psi_\mu, Z_\mu \} + 6 \,\{\psi_\mu, 2 \,Z_\mu, \chi \} + 14 \,\{ Z_\mu, 2\,\chi, \varphi\} +
14 \,\{ \chi, 2\,\varphi\}
\end{align}
In the case of a ${\bf Z}_3$ orbifold, whereby $x^i \rightarrow z^I, z^*_{\bar{I}}$ the multiplicities can be expressed in terms of dimensions of irreps of $SU(3)$ \ie{} ${\bf 6} \rightarrow {\bf 3} + {\bf 3^*}$,
${\bf 14} \rightarrow {\bf 8} + {\bf 3} + {\bf 3^*}$ and ${\bf 14'} \rightarrow {\bf 1}+ {\bf 1} + {\bf 6} + {\bf 6^*}$.
Once again, it is not surprising that the multiplet content $\{H_{\mu\nu}, 2 \psi_\mu, Z_\mu \}$ be in one-to-one correspondence with the currents  $\{T_{\mu\nu}, \Sigma_\mu, \bar\Sigma_\mu, J_\mu \}$ in the $\cN = 1$ super-current multiplet of Ferrara and Zumino \cite{Ferrara:1974pz}. 

For later purposes, note that $H_{\mu\nu}$, with $\eta^{\mu\nu} H_{\mu\nu} = 0 = p^\mu H_{\mu\nu}$ belongs in a spin-2 supermultiplets with 8 bosonic and as many fermionic d.o.f. whose vector boson is $Z_\mu = \delta^{I\bar{J}} C_{\mu I \bar{J}}$, while $H'_{\mu\nu} = H_0 [\eta_{\mu\nu} + \ap p_\mu p_\nu] $ with $H_{ij} = H_0 \delta_{ij}/2$ (so that $\eta^{MN} H_{MN} = 0$) combine with ${C}_0=\varepsilon_{\mu\nu\rho\sigma}p^\sigma{C}^{\mu\nu\rho}/M$ in a chiral  multiplet.

\subsection{Four-point amplitudes (superstring) and spinor helicity basis}

For simplicity, we will only consider amplitudes with a single massive external state: $\cA({A}_1,{A}_2,{A}_3,{H}_4)$ and $\cA({A}_1,{A}_2,{A}_3,{C}_4)$. Depending on the choice of incoming particles these correspond to production, annihilation or 3-body decay of the massive state. In view of this, it is useful to restrict to 4-dimensional momenta and polarisations and work in the helicity basis whereby null momenta are expressed in terms of on-shell Weyl spinors of opposite chirality
\be
k_{\alpha\dot\alpha} = k_\mu\sigma^\mu_{\alpha\dot\alpha} = u_\alpha \bar{u}_{\dot\alpha}
\ee
and resort to the standard notation $u_\alpha (k) \rightarrow |k\rangle$  $\bar{u}_{\dot\alpha}(k) \rightarrow [k|$, so much so that 
\be
u(k_i) u(k_j) = - \langle i  j\rangle \quad , \quad \bar{u}(k_i) \bar{u}(k_j) = [ij] \quad {\rm and} \quad 2\,k_i{\cdot}k_j = \langle i j\rangle [ij]
\ee
For real momenta $\bar{u}_{\dot\alpha}(k) = (u_\alpha(k))^*$. Momentum conservation reads
\be
\sum_i |i\rangle [i| = 0 = \sum_i |i] \langle i|.
\ee
Schoutens's identity entails $\langle 12\rangle \langle 34\rangle {+} \langle 13\rangle \langle 42\rangle {+} \langle 14\rangle \langle 23\rangle{ =} 0$ and a similarly for $|k]$'s. 

Positive and negative helicity polarisations can be expressed as 
\be
a_{\alpha\dot\alpha}^- = a^-_\mu\sigma^\mu_{\alpha\dot\alpha} = { u_\alpha \bar{v}_{\dot\alpha} \over \bar{u} \bar{v}}
\qquad
{\rm and}
\qquad
a_{\alpha\dot\alpha}^+ = a^+_\mu\sigma^\mu_{\alpha\dot\alpha} = { v_\alpha \bar{u}_{\dot\alpha} \over {u} {v}},
\ee
where $q_{\alpha\dot\alpha} = v_\alpha \bar{v}_{\dot\alpha}$ is an arbitrary light-like momentum that encodes the gauge freedom.   

Also for massive particles it proves convenient to express their momenta and polarisations in terms of null momenta and Weyl spinors. Setting $p_{\al\ald} {=} k_{\alpha\dot\alpha} {+} q_{\alpha\dot\alpha}{=}u_{\al}\bar u_{\ald}{+}v_\alpha \bar{v}_{\dot\alpha}$ one has $p^2 {=} 2kq{=}{-}m^2{=}uv\,\bar u\bar v$.

Helicity of a massive particle is not Lorentz invariant. For later purposes it proves convenient to explicitly identify the precise Lorentz transformations that map massive helicity states into one another. Let us choose the basis $\{u_\alpha, v_\alpha\}$ for Left-handed spinors with $uv {=} \langle u v \rangle {\neq} 0$ and $\{\bar u_{\dot\alpha}, \bar v_{\dot\alpha}\}$ for Right-handed spinors with $\bar u \bar v {=} [u v] {\neq} 0$. Dropping indices for simplicity, the Lorentz group $SL(2, C) \times \overline{SL(2, C)}$ act as
\be
L u = u' = { a u + b v} \quad  L v = v' = { c u + d v }  
\ee
with $a,b,c,d \in {\bf C}$ such that $ad-bc = 1$ and 
\be
R \bar{u} = \bar{u}' = { \bar{a} \bar{u} + \bar{b} \bar{v} } \quad  
R \bar{v} = \bar{v}' = { \bar{c} \bar{u} + \bar{d} \bar{v} }
\ee
It is easy to check that any symplectic product is invariant \ie$ \langle i'j'\rangle {=} \langle ij\rangle$ and  $ [i'j'] {=} [ij]$. The Lorentz transformations that leave the time-like momentum $p$ invariant form an $SO(3)$ subgroup with
\be
a = e^{i\alpha} \cos{\gamma} \quad , \quad b = e^{i\beta} \sin\gamma  \quad , \quad 
c = - e^{-i\beta} \sin\gamma \quad , \quad d = e^{- i\alpha}\cos\gamma 
\ee
The $SO(3)$ transformations 
\be
L_x \quad : \quad u' = {1\over \sqrt{2}} (u+v) \quad v' = {1\over \sqrt{2}} (- u + v)
\ee
and 
\be
L_y \quad : \quad u' = {1\over \sqrt{2}} (u + i v) \quad v' = {1\over \sqrt{2}} (i u + v)
\ee
with $R_{x/y}{=}L_{x/y}^*$ will prove particularly useful in the following.

For a massive vector boson, with $p {=} u \bar{u} {+} v \bar{v}$  the three helicity states are\footnote{For a different basis of massive polarisations in 4-pt amplitudes, see \cite{Bianchi:2010dy,Bianchi:2010es}.}  
\be
w_0 = u \bar{u} - v \bar{v} \quad w_+ = u \bar{v} \quad w_- = v \bar{u}
\ee
with $w_0{\cdot} w_0 {=} 4 m^2$, $w_0{\cdot} w_{\pm}{ =} 0$, $w_\pm{\cdot} w_{\pm} {=} 0$,
$w_\pm{\cdot} w_{\mp} {=} 2m^2$. $\{w_0, w_+, w_-\}$ form a complete basis for transverse polarisations in that
\be
w_0{\otimes}w_0 +  w_+{\otimes}w_- + w_-{\otimes}w_+ = 2 m^2 \eta + 2 p{\otimes}p
\ee
The complex circular polarisations $w_{\pm}$  can be combined into real ones
\be
w_x = u \bar{v} + v \bar{u} \quad w_y = i u \bar{v} - i v \bar{u}
\ee
It is easy to check that $L_x$ maps $w_x$ into $w_0$ (up to a sign $L_x w_x {=} {-} w_0$) and  {\it vice versa} $L_x w_0 {=} {+} w_x$, leaving $w_y$ unaltered $L_x w_y {=} w_y$, while $L_y$ maps $w_y$ into $w_0$ ($L_y w_x {=} w_0$) and {\it vice versa} $L_y w_0 {=}{ -} w_y$, leaving $w_x$ unchanged $L_y w_x {=} w_x$.

For a massive tensor boson ($s{=}2$), the five helicity states can be taken to be 
\be
H_{++} = w_+{\otimes} w_+ \quad   H_{--} = w_-{\otimes} w_-  \quad H_{00} = w_0{\otimes} w_0 - w_+{\otimes} w_- - w_-{\otimes} w_+
\nonumber\ee
\be
H_{+0} = w_+{\otimes} w_0 + w_0{\otimes} w_+ \quad  H_{-0} = w_-{\otimes} w_0 + 
w_0{\otimes} w_- \ee
Note that $w_0{\otimes}w_0 +  w_+{\otimes}w_- + w_-{\otimes}w_+ = 2 m^2 \eta + 2 p{\otimes}p
$ is a scalar polarisation. 
As for the vector polarisations, the complex combinations $H_{\pm\pm}$ and $H_{\pm 0}$ can be combined into real ones $H_{xx}-H_{yy} {=} H_{++} + H_{--}$, $H_{xy} {=} i H_{++} - iH_{--}$, $H_{x0} {=} H_{+0} {+} H_{-0}$, $H_{y0} {=} i H_{+0} {-} iH_{-0}$ ($H_{00}$ is real).
The transformation $L_x{+}L_y$ leaves $H_{++}{+}H_{--}$ invariant, while $L_x{-}L_y$ maps $H_{++}{+}H_{--}$ into $H_{00}$. $L_x$ maps $H_{++}{-}H_{--}$ into $(H_{-0}{-}H_{+0})/2$ while 
$L_y$ maps $H_{++}{-}H_{--}$ into ${-}i(H_{+0}{+}H_{-0})/2$.

For spin $s$ totally symmetric tensors (as in the first Regge trajectory) one has $2s{+}1$ helicity states, starting from the `top' component $S_{++\ldots{+}} {=} (u\bar{v})^s=w_{+}^s$ to the `bottom' $S_{--\ldots{-}} {=} (v\bar{u})^s {=} w_{-}^s$, passing through the middle components $S_{00\ldots{0}} {=} (u\bar{u} {-} v\bar{v})^s {+} \dots {=} w_0^s + \dots$. Applying combinations of the above $SO(3)$ transformations (on the helicity spinors) one can map any amplitude, \eg the one with the `top' helicity component of a massive state, into any other. This applies independently for each external insertion.

\subsubsection{Amplitude $\cA_{{A}{A}{A}{C}}$ in $D{\le} 10$}

Let us start with $\cA({A}_1,{A}_2,{A}_3,{C}_4)$.
With a judicious choice of super-ghost pictures and c-ghost insertions one has
\begin{align}
&\cA\left({A}^{(-1)}_1,{A}^{(0)}_2,{A}^{(0)}_3,{C}^{(-1)}_4\right) {=} \lim_{(z_1,z_2,z_4) \rightarrow (\infty, 1, 0)} \int_{0}^{1} dz_3 \bigr\langle c e^{-\varphi} a_1\psi e^{ik_1X}(z_1) \nonumber\\
&c (a_2 i\de X {+} k_2\psi a_2\psi)e^{ik_2X}(z_2)
(a_3 i\de X {+}  k_3\psi a_3\psi)e^{ik_3X}(z_3) c C_4 \psi\psi\psi e^{ip_4X}(z_4)\bigr\rangle
\end{align}
Following the steps detailed in Appendix~\ref{app:AAAC}, one finally gets 
\begin{align}
\cA_{{A}{A}{A}{C}} {=}  \cB (1,1) &\biggr\{ {-} a_1{\cdot}C_4{:}f_2 \left[ a_3{\cdot}k_1 {-} {u \over t} {a_3{\cdot} k_2} \right]
{-}  a_1{\cdot}C_4{:}f_3 \left[ {u \over t} {a_2{\cdot}k_3} {-} {u \over s} {a_2{\cdot} k_1}\right]
{+} {u \over s} {a_1{\cdot}f_2{\cdot}C_4{:}f_3 }  \nonumber\\
&{-}  {a_1{\cdot}f_3{\cdot} C_4{:}f_2} 
{-} 2 {u \over t} {a_1{\cdot}\dot{C}_4\overline{{\cdot}f_3{\cdot}}\dot{f}_2 } \biggr\},
\label{eq:AAAC}
\end{align}
where the contractions are performed in a self-explanatory fashion and 
\be
\cB (1,1) {=} \cB(2\ap k_3p_4, 1 {+} 2\ap k_2k_3) {=} \frac{\Gamma(2\ap k_3p_4)\Gamma(1{+}2\ap k_2k_3)}{\Gamma(1{+}2\ap k_3(k_2+p_4))}.
\ee
Using $2k_1k_2 {=} {-}s {=} {+}2k_3p_4 {-} M^2$, $2k_2k_3 {=} {-}t {=} {+}2k_1p_4 {-} M^2$ and $2k_3k_1 {=} {-}u {=} {+}2k_2p_4 {-} M^2$ \emph{i.e.} 
$s{+}t{+}u{ =} M^2 {=} 1/\ap$, one can check gauge invariance with respect to each of the three vector legs. 

Expanding and shuffling all the terms in Eq.~\eqref{eq:AAAC}, the amplitude $\cA_{{A}{A}{A}{C}}$ can be written in a manifestly symmetric form under the exchange of the three 
vector boson legs
\begin{align}
\cA_{{A}{A}{A}{C}}&=4\alp u\,\cB(1,1)\biggr(C_4[a_1a_2a_3]+\sum_{i\neq 3} C_4[a_1a_2k_i]\frac{a_3k_i}{k_3k_i}+\sum_{i\neq 2} C_4[a_3a_1k_i]\frac{a_2k_i}{k_2k_i}+\nonumber\\
&\sum_{i\neq 1} C_4[a_2a_3k_i]\frac{a_1k_i}{k_1k_i}+C_4[a_1k_2k_3]\frac{a_2a_3}{k_2k_3}+C_4[a_2k_3k_1]\frac{a_3a_1}{k_3k_1}+C_4[a_3k_1k_2]\frac{a_1a_2}{k_1k_2}\biggr),
\end{align}
where $C[abc]{=}C_{\mu\nu\rho}a^\mu b^\nu c^\rho$.
\subsubsection{$\cA_{{A}{A}{A}{C}}$ in $D{=}4$ helicity basis}
In $D{=}4$ $C_{\mu\nu\rho}$ is equivalent to a (pseudo) scalar $C_0{=} \varepsilon^{\lambda\mu\nu\rho}p_\lambda C_{\mu\nu\rho}/6M$. In the helicity basis one has two independent color-ordered amplitudes $\Amp(1^-2^-3^-{C}_0)$ and 
$\Amp(1^-2^-3^+{C}_0)$ and their complex conjugates $\Amp(1^+2^+3^+{C}_0)$ and $\Amp(1^+2^+3^-{C}_0)$\footnote{Once again, details of the computations are relegated in Appendix~\ref{AAAC_4D}.}.
The former reads
\be 
\Amp(1^-2^-3^-{C}_0)=i\cB(1,1)\frac{ m_{C} \lef 13 \re}{[12][23]}.
\ee
The latter reads
\be
\Amp(1^-2^-3^+{C}_0) = i\cB(1,1)\frac{[13] \lef 12\re^3}{ m_{C}^3 \lef 23\re}.
\ee

\subsubsection{Amplitude $\cA_{{A}{A}{A}{H}}$ in $D {\le} 10$}
Let us now consider $\cA_{{A}{A}{A}{H}} $. 
With a judicious choice of super-ghost pictures and c-ghost insertions one has
\begin{align}
&\cA\biggr(A^{(0)}_1,A^{(-1)}_2,A^{(0)}_3,H^{(-1)}_4\biggr) {=} \lim_{(z_1,z_2,z_4) \rightarrow (\infty, 1, 0)} \int_0^1 dz_3
\bigr\langle c (a_1 i\de X {+}  k_1\psi a_1\psi)e^{ik_1X}(z_1) \nonumber\\
&c e^{-\varphi} a_2\psi e^{ik_2X}(z_2) \,
(a_3 i\de X {+} k_3\psi a_3\psi)e^{ik_3X}(z_3)\, c \de X{\cdot}H_4{\cdot}\psi e^{ip_4X}(z_4)\bigr\rangle.
\end{align}
Following the steps detailed in Appendix~\ref{app:AAAH}, one finally finds 
\begin{align}
&\cA_{{A} {A} {A} {H}} {=}\frac{1}{st}\cB(1{{-}}\alp s,1{{-}}\alp t)\{ a_1a_3\,st[a_2Hk_1(1{{-}}\alp s){{-}}a_2Hk_3(1{{-}}\alp t)] \nonumber\\
&{+} 2a_1Ha_2\,s\,k_1f_3k_2{-}2a_3Ha_2\,t\,k_3f_1k_2{+}2\alp[st(a_3k_2\,a_1k_3\,k_3Ha_2{-}a_1k_2\,a_3k_1\,k_1Ha_2)\nonumber\\
&{+}a_1k_2\,a_3k_2\,u\,(k_1Ha_2\,s{-}k_3Ha_2\,t)]{-}2a_1k_2\,a_3k_1\,a_2Hk_3(1{-}\alp t)t{+}2a_3k_2\,a_1k_3\,a_2Hk_1(1{-}\alp s)s\nonumber\\
&{+}a_2f_3Ha_1\,us{-}a_2f_1Ha_3\,ut{-}2\alp(a_3k_1\,a_2f_1Hk_1{-}a_1k_3\,a_2f_3Hk_3)st\nonumber\\
&{+}2\alp a_1k_2(a_2f_3Hk_1\,s{-}a_2f_1Hk_3\,t)u{-}2\alp a_3k_2(a_2f_1Hk_3\,t {-}a_2f_1Hk_1\,s)u\nonumber\\
&{+}2a_1k_3\,a_2f_3Hk_1(1{-}\alp s)s{-}2a_3k_1\,a_2f_1Hk_3(1{-}\alp t)t{-}2\alp st(a_2f_1f_3Hk_1{-}a_2f_3f_1Hk_3)\nonumber\\
&{-}2a_2f_1f_3Hk_3(1{-}\alp t)t{+}2a_2f_3f_1Hk_1(1{-}\alp s)s\}.
\end{align}
After laborious manipulations, this amplitude can be written in the compact symmetric form
\begin{align}
\cA_{{A} {A} {A} {H}} {=}\frac{4}{st}\cB(1{-}\alp s,1{-}\alp t)&\{\left[2\alp f_3Hf_1\,k_3f_2k_1{+}(1{+}2\alp k_1k_3)f_3Hf_1f_2\right]k_3k_1\nonumber\\
&{+}\left[2\alp f_1Hf_2\,k_1f_3k_2{+}(1{+}2\alp k_1k_2)f_1Hf_2f_3\right]k_1k_2\nonumber\\
&{+}\left[2\alp f_2Hf_3\,k_2f_1k_3{+}(1{+}2\alp k_2k_3)f_2Hf_3f_1\right]k_2k_3\}.
\label{eq:AAAH}
\end{align}

\subsubsection{$\cA_{{A}{A}{A}{H}}$ in $D{=}4$ helicity basis}

Let us first consider the amplitudes involving the scalar component of ${H}$ and start with $\cA(1^+2^+3^+{H}_0) {=} \cA(1^{-}2^{-}3^{-}{H}_0)^*$.
The amplitude can be written in the very compact form
\be \cA(1^+2^+3^+{H}_0) =  \,\cB(1,1) \frac{ m_{H} [13]}{\lef 12 \re \lef 23\re}. \ee
which is identical up to a phase to $\cA(1^+2^+3^+{C}_0)$, for normalised states.

Consider a different choice for the helicity of the vectors in the amplitude with ${H}_0$: $\cA(1^-2^-3^+{H}_0) {=} \cA(1^+2^+3^-{H}_0)^*$.
The final result reads 
\be \Amp(1^-2^-3^+{H}_0)=\,\cB(1,1) \frac{[13] \lef 12\re^3 }{ m_{H}^3 \lef 23\re}. \ee
which is identical up to a phase to $\cA(1^-2^-3^+{C}_0)$.

Consider now the the amplitude for the spin-2 tensor ${H}_2$ and three vector bosons  
$\cA(1^-2^+3^+{H}^h_2) {=} \cA(1^+2^-3^-{H}^{{-}h}_2)^*$. Setting $p{=}k_4{+}k_5$, the  simplest amplitude to compute is the one for the state with polarisation ${H}^{++} {=} \lef 4| \lef 4| |5] |5]$ 
that reads 
\be \cA(1^-2^+3^+H^{++}) = 
 \cB(1,1) \frac{\lef 14 \re^4 [13]}{m_H \lef 12 \re \lef 23 \re \lef 45 \re^2}.
\ee
The other amplitudes obtain in a straightforward way, after repeatedly applying $L_x$ and $L_y$ as outlined above. The final result can be compactly written as
\begin{align} &\sum_h c_h \cA(1^-2^+3^+H^{h}) =  \nonumber\\
& \cB(1,1) \frac{ [13] \lef 14 \re^2 \lef 15 \re^2}{m_H \lef 12 \re \lef 23 \re \lef 45 \re^2} \left\{ c_{++} {\lef 14 \re^2 \over \lef 15 \re^2} - 4 c_{+0} {\lef 14 \re \over \lef 15 \re}  + 6 c_{00}  - 4 c_{0-} {\lef 15 \re \over \lef 14 \re} + c_{--} {\lef 15 \re^2 \over \lef 14 \re^2}.
\right\}
\end{align}
In the chosen orthogonal basis $|H_{++}|^2 {=} |H_{--}|^2 {=} 4(k_4k_5)^2{=}m_H^4 {=}(1/\ap)^2$, 
$|H_{+0}|^2 {= }|H_{0-}|^2 = 16(k_4k_5)^2{=}4m_H^4 {=}(2/\ap)^2$  and $|H_{00}|^2 {=} 24(k_4k_5)^2{=}6m_H^4 {=}(\sqrt{6}/\ap)^2$, so much so that $\hat{c}_{\pm\pm} = m_H^2 {c}_{\pm\pm} $, $\hat{c}_{\pm 0} {=} 2 m_H^2 {c}_{\pm 0} $  and $\hat{c}_{00} {=} \sqrt{6} m_H^2 {c}_{00} $ for properly normalized polarization tensors.

\subsection{Higher-point open superstring amplitudes from SYM}
In \cite{Mafra:2011nv, Mafra:2011nw}  Mafra, Schlotterer and Stieberger (MSS) have obtained a beautiful formula that allows one to express open superstring amplitudes for massless external states on the disk to SYM amplitudes at tree level. The formula is reminiscent of the KLT (Kawai, Lewellen, Tye) relations \cite{Kawai:1985xq} and reads  
\be 
\cA^{ST}_n(1,2,\ldots n) =\sum_{\sigma\in S_{n-3}} F(1[2_\sigma 3_\sigma \ldots n~-~2_\sigma]n-1,n) \cA^{YM}_n(1[2_\sigma 3_\sigma \ldots n-2_\sigma]n-1,n)  
\ee
with $z_1=0,z_{n-1}=1,z_n=\infty$ so that
\be
F(1[23\ldots n-2]n-1,n) = \nonumber \\
\ee
\be
\quad (-)^{n-3} \int^1_{z_1=0} dz_2 \int^1_{z_2} dz_3 \ldots \int^{z_{n-1}=1}_{z_{n-3}} dz_{n-2}
\prod_{i<j} z_{ij}^{s_{ij}} \prod_{k=2}^{[n/2]} \sum_{l=1}^{k-1} {s_{lk} \over z_{lk}} \prod_{k=[n/2] + 1}^{n-1} \sum_{l=k+1}^{k-1} {s_{kl} \over z_{kl}},
\label{eq:MSST}
\ee
where $s_{ij} {=} 2\ap k_ik_j {=} {-}\ap s_{ij}^{phys}$. 
The formula \eqref{eq:MSST} follows from a tree-level CFT computation using the pure spinor formalism \cite{Mafra:2011nv} and its soft limits and other properties were checked in \cite{Mafra:2011nw}. A pure RNS derivation of Eq.~\eqref{eq:MSST} has been given in~\cite{Barreiro:2013dpa}, the proof is based on a revisited S-matrix approach~\cite{Barreiro:2012aw} .We will here check that it is consistent with factorization on massive string states in two-particle channels \emph{i.e.}
\be 
\lim_{s_{12} \rightarrow - \ap M_H^2} (s_{12} + \ap M_H^2) \cA_{n}(V_1V_2V_3\ldots V_n)  = \sum_H \cA_3(V_1V_2H) \cA_{n-1}(HV_3V_4\ldots V_n),
\ee
where  $\cA_3(VVH)$ is physical (decay rate, width) and can be computed for arbitrary states following the strategy outlined in appendix~\ref{app:supervertex}. This is nothing but $\text{Res}[\cA_{n}(V_1V_2V_3\ldots V_n)]$ for $s_{12} {=} {-} \ap M_H^2$.

For simplicity will only consider mass-less 5-point amplitudes producing 4-point amplitudes with 3-massless and 1-massive state in $D{=}4$ 
and briefly mention how to generalize the procedure to an arbitrary number of mass-less and massive external  states. In particular we give the relevant formula for mass-less 6-point amplitudes and sketch, at least in the MHV case, how to get the 4-massless and 1-massive at 5-points or the 2-massless and 2-massive at 4-points.

\subsection{5-points in $D{=}4$ helicity basis}
The 5-point color-ordered amplitude for open superstring massless gluons reads \cite{Mafra:2011nv, Mafra:2011nw}
\be 
\cA_5(12345) = F(12345) \cA^{YM}_5(12345) + F(13245) \cA^{YM}_5(13245),
\ee
where $F$ are multiple hyper-geometric functions
\be
F(1[23]45) = s_{12} s_{34} \int_0^1 dx \int_x^1 dy x^{s_{12} -1} y^{s_{13}} (y-x)^{s_{23}}
(1-x)^{s_{24}} (1-y)^{s_{34} - 1},
\ee
with $s_{ij} {=} 2\ap k_ik_j$ and $F(13245)$ is obtained by exchanging 2 and 3 \emph{i.e.}
\be
F(1[32]45) = s_{13} s_{24} \int_0^1 dx \int_x^1 dy x^{s_{12}} y^{s_{13}-1} (y-x)^{s_{23}} 
(1-x)^{s_{24}-1} (1-y)^{s_{34}}.
\ee
Since in $D{=}4$ any 5-pt amplitude is either MHV or antiMHV with $\cA^{\overline{MHV}}_5(1^+2^+3^-4^-5^-)= {\cA^{MHV}_5}(1^-2^-3^+4^+5^+)^*$, let us consider the MHV case for definiteness 
\be 
\cA_5(1^-2^-3^+4^+5^+) = {\langle 12\rangle^3\over \langle 23\rangle \langle 34\rangle \langle 45\rangle \langle 51\rangle} F(12345)  + {\langle 12\rangle^4\over \langle 13\rangle \langle 32\rangle \langle 34\rangle \langle 45\rangle \langle 51\rangle}  F(13245)  
\ee
that can be written as 
\be
\cA_5(1^-2^-3^+4^+5^+) = {\langle 12\rangle^3\over \langle 13\rangle \langle 23\rangle \langle 34\rangle \langle 45\rangle \langle 51\rangle} [  \langle 13\rangle F(12345) - \langle 12\rangle
F(13245)].
\ee
MSS have checked the correct factorization on the massless poles \cite{Mafra:2011nv, Mafra:2011nw}.
Here we will check consistency in the massive two-particle channel. To this end one has to
take the residue at the pole $s_{12}\to {-}1$ respectively of $F(12345)$ and $F(13245)$.
Starting from the expression 
 \begin{align}
F(12345)=s_{12}s_{34}\int_0^1 dy \int_0^y  dx \,x^{s_{12}-1}(1-x)^{s_{24}}(y-x)^{s_{23}}(1-y)^{s_{34}-1}y^{s_{13}}
\label{eq:F12345}
\end{align}
and making use of
 \be
x^{s_{12}-1}=\frac{1}{s_{12}(s_{12}+1)}\frac{d^2}{dx^2}x^{s_{12}+1}
 \ee
 in Eq.~(\ref{eq:F12345}) and integrating by parts, one finds
 \begin{align}
F(12345)=s_{34}\int_0^1 dy  \, (1-y)^{s_{34}-1}y^{s_{13}} \int_0^y  dx \, \frac{x^{s_{12}+1}}{s_{12}+1}  \frac{d^2}{dx^2}[(1-x)^{s_{24}}(y-x)^{s_{23}}].
 \end{align}
 Now it is easy to take the residue and find
 \begin{align}
\underset{s_{12} = -1}{\text{Res}}  F(12345) & = \lim_{s_{12}\to-1}(s_{12}+1)F(12345) \nonumber \\ & = s_{34}s_{24}\cB(s_{13}+s_{23}+1,s_{34})+s_{34}s_{23}\cB(s_{13}+s_{23},s_{34}).
 \end{align} 
Performing the same steps for 
\be
F(13245)=s_{13}s_{24}\int_0^1 dy \int_0^y  dx \,x^{s_{12}}(1-x)^{s_{24}-1}(y-x)^{s_{23}}(1-y)^{s_{34}}y^{s_{13}-1},
\ee
yields
 \begin{align}
\underset{s_{12} = -1}{\text{Res}} F(13245)= \lim_{s_{12}\to-1}(s_{12}+1)F(13245) =s_{13}s_{24}\cB(s_{13}+s_{23},s_{34}+1).
 \end{align}
 Finally the residue of the color-ordered string amplitude is
 \begin{align}
&\underset{s_{12} = -1}{\text{Res}} \Amp_5^{ST}(12345) = s_{34}\cB(s_{3p},s_{34})\biggr\{\Amp_5^{YM}(12345)\left[s_{23}-\frac{s_{24}s_{3p}}{s_{35}} \right] -\Amp_5^{YM}(13245)\frac{s_{13}s_{24}}{s_{35}}\biggr\},
\label{eq:Amp5st}
 \end{align}
 where $s_{3p}{=}s_{13}{+}s_{23}{=}{-}s_{34}{-}s_{35}{=}\alp t{+} \alp u{=}1{-} \alp s$, since $p{=}k_1{+}k_2$ and $2k_1k_2{=}p^2{=}{-}M^2_{H/C}$.
Using a mixed notation with both physical Mandelstam variables ($s,t,u$) and $s_{ij}$ variables we obtain the following expression
\begin{align}
&\underset{s_{12} {=} {-}1}{\text{Res}} \Amp_5^{ST}(12345)\nonumber\\
&{=}\cB(1{-}\alp s,1{-}\alp t)\bigr\{ \Amp_5^{YM}(12345)[ s_{23}s_{35}{+}(s_{34}{+}s_{35})s_{24}] {-}s_{13}s_{24}\Amp_5^{YM}(13245) \bigr\}.
\end{align}
One can check the factorization case by case, fixing the helicity of the external gluons. Before embarking in the computations, notice that only $SO(6)$ singlet bosons can appear in the two-gluon channel. Following the dimensional reduction we previously revisited in some detail, one only has $H_{\mu\nu}, C_{\mu\nu\rho} {=} C_0 \varepsilon_{\lambda\mu\nu\rho} p^\lambda/m_{{C}} $ and $\delta^{ij} H_{ij} {=} {-} \eta^{\mu\nu} H_{\mu\nu} {=} {-} 3 H_0$, after decomposing
$ H_{\mu\nu} {=} H^{tt}_{\mu\nu} {+} H_0 (\eta_{\mu\nu} { +}\ap p_\mu p_\nu)$.
Let us start with $\Amp(1^-2^-3^+4^+5^+)$. In this case we expect that only ${C}$ and $3 {H}_0{=}\eta^{\mu\nu}{H}^{(4)}_{\mu\nu} {=} \eta^{\mu\nu}{H}^{(10)}_{\mu\nu} {=} {-} \delta^{ij} {H}^{(10)}_{ij}$, with ${H}^{(4)}_{\mu\nu} {=} \eta_{\mu\nu} {+} \ap p_\mu p_\nu$ the four-dimensional part of ${H}$, contribute. With this choice, Eq.~(\ref{eq:Amp5st}) becomes
 \begin{align}
&\underset{s_{12} = -1}{\text{Res}} \Amp_5^{ST}(1^-2^-3^+4^+5^+)\nonumber\\
&=\cB(1,1)\frac{\lef 12 \re^4}{\lef 23 \re \lef 45 \re \lef 51 \re} \biggr[ \frac{\lef 23 \re [23] \lef 35 \re [35] + (\lef 34 \re [34]+\lef 35 \re [35])\lef 24 \re [24]}{\lef 12 \re \lef 34 \re}-[13][24]\biggr]\nonumber\\
&{=} { \lef 12 \re^2 \over m_{H/C}}\cB(1,1)\frac{m_{H/C} [35]}{\lef 45 \re \lef 34 \re} {=} \cA_3(1^-2^-{H}_0) \cA_4({H}_03^+4^+5^+){+}\cA_3(1^-2^-{C}_0) \cA_4({C}_03^+4^+5^+),
 \end{align}
 where $\cB(1,1){=}\cB(1{+}2\alp k_1k_2,1{+}2\alp k_1k_3)$. The result coincides with the one we previously derived using standard world-sheet techniques.

  Consider now the amplitude $\cA(1^-2^+3^-4^+5^+)$. 
 As shown in Tab.~\ref{tab:VVH} in appendix~\ref{AAAH_D4}, if we take $k_{1\al \ald}{=}u_\al \bar u_{\ald}$ and $k_{2\beta \betad}{=}v_\beta \bar v_{\betad}$, with $k_1$, $k_2$ such that $2k_1\cdot k_2{=}p^2{=}{-}1/\alp$, we find 
 that only the spin-2 polarization $v_\al v_\beta \bar u_{\ald} \bar u_{\betad}$ contributes at the massive pole.
 With this choice, we have
 \begin{align}
&\underset{s_{12} = -1}{\text{Res}} \Amp_5^{ST}(1^-2^+3^-4^+5^+)\nonumber\\
&=\cB(1,1)\frac{\lef 13 \re^4 \{\lef 23 \re [23]\lef 35 \re [35]+(\lef 34 \re [34]+\lef 35 \re [35])\lef 24 \re [24]+[13][24]\lef 12 \re \lef 34 \re\}}{\lef 12 \re \lef 23 \re \lef 34 \re \lef 45 \re \lef 51 \re}  \nonumber\\
&= m_H \cB(1,1)\frac{\lef 13 \re^4[35]}{m_H \lef 12 \re^2 \lef 34 \re \lef 45 \re } = \sum_h \cA_3(1^-2^+H^{h}) \cA_4(H^{-h} 3^-4^+5^+).
 \end{align}
 where only $H^{++} {=} |1\re |1\re |2] |2]$ contributes since $\cA_3(1^-2^+H^{++}) {=} m_H {=} {-} 2 \sqrt{\ap} k_1k_2$, while 
 for the remaining helicity states $\cA_3(1^-2^+H^{h\neq ++}) {=} 0$ . 
 The result coincides with the one we previously derived using standard world-sheet techniques.
   
The last case is the amplitude $\cA(1^+2^+3^-4^-5^+)$ in which, as for the first case, only $H_0$ and $C_0$ get exchanged in the $s_{12}$ channel.
\begin{align}
&\underset{s_{12} = -1}{\text{Res}} \Amp_5^{ST}(1^+2^+3^-4^-5^+)\nonumber\\
&=\cB(1,1)\frac{\lef 34 \re^3}{\lef 12 \re \lef 23 \re \lef 45 \re \lef 51 \re}(\lef 23 \re [23]\lef 35 \re [35]+(\lef 34 \re [34]{+}\lef 35 \re [35])\lef 24 \re [24]{+}[13][24]\lef 12 \re \lef 34 \re)\nonumber\\
&= {[12]^2 \over m_{H/C}} \cB(1,1)\frac{\lef 34 \re^3 [35]}{m_{H/C}^3 \lef 45 \re} = \cA_3(1^+2^+{H/C}) \cA_4({H/C},3^-4^-5^+).
\end{align}
The result coincides with the one previously derived using standard world-sheet techniques.
  
 \subsection{6-points and higher point amplitudes}
 
Open-string amplitudes with more than one massive insertion look somewhat cumbersome and not very illuminating in $D{=}10$. In $D{=}4$, in the spinor helicity basis, formulae look more compact. A possible strategy for systematic computations is to derive amplitudes for massive states by multiple factorization of amplitudes for massless states on massive poles in two-particle channels. For open superstrings in turn one can rely on the MSS formula \cite{Mafra:2011nv, Mafra:2011nw}, relating string amplitudes to SYM amplitudes, whose validity we have given further support earlier on.   

For instance at 6-points, there are six terms in the MSS formula, corresponding to the permutations of [234] \ie
\begin{align} 
\cA_6^{ST}(123456) &= F(1[234]56) \cA_6^{YM}(1[234]56) + F(1[342]56) \cA_6^{YM}(1[342]56) \nonumber\\ 
&+ F(1[423]56) \cA_6^{YM}(1[423]56) + F(1[324]56) \cA_6^{YM}(1[324]56) \nonumber\\
&+ F(1[432]56) \cA_6^{YM}(1[432]56) + F(1[243]56) \cA_6^{YM}(1[243]56)
\end{align}
Differently from the 4- and 5-point cases where only MHV (or anti-MHV) amplitudes are non-zero, at 6-point one has a NMHV amplitude $\cA^{NMHV}_6(---+++)$ that even in SYM has a lengthy expression if compared to Parke-Taylor formula \cite{Mangano:1990by}. Focussing on MHV amplitudes $\cA_6^{MHV}(--++++) {=} \lef 12 \re^3/\lef 23 \re \lef 34 \re \ldots \lef 61 \re$ one can still compute 5-point amplitudes with one massive insertion with almost no effort and 4-point amplitudes with two massive insertions with little more effort. 

For an arbitrary number of external massless legs $n$ {\it a priori} one has N$^k$MHV amplitudes with $k=0, \ldots [n/2]{-}2$. These, and susy related ones, are needed to compute amplitudes for generic massive states by factorization. Summarizing one can start with $\cA^{SYM, k}_{n+2m}$, then derive $\cA^{ST, k}_{n+2m}$ and finally obtain $\cA^{ST, k}_{n,m}$ by factorization on the assigned two-particle massive poles. Notice that the initial helicity configuration should be chosen compatibly with the choice of massive states, \emph{i.e.} at the first level $H^0/C^0$ couple to gluons with the same helicity while $H_2$ couples to gluons with opposite helicity. Reverting the argument, the allowed helicity configurations in SYM constrain the allowed amplitudes in superstring theory.

\section{Soft limit}

\subsection{General arguments}

In \cite{Bianchi:2014gla} the soft limits of massless string amplitudes was studied both explicitly (up to 6-point amplitudes) and abstractly by making use of OPE analysis. The conclusion was that disk amplitudes of gluons behave exactly as in Yang-Mills theory at tree level both for the open superstring and for the open bosonic string. 
Indeed one expects universal behaviour at leading ($\delta^{-1}$) and sub-leading ($\delta^{0}$) order, in formulae 
\be
\cA_{n}(1,2,\ldots,s,\ldots,n) =  \nonumber\ee
\be \left\{ \left[{a_s{\cdot}k_{s+1} \over k_s{\cdot}k_{s+1}} - {a_s{\cdot}k_{s-1} \over k_s{\cdot}k_{s-1}} \right] + \left[{f_s{:}J_{s+1} \over 2 k_s{\cdot}k_{s+1}} - {f_s{:}J_{s-1} \over 2 k_s{\cdot}k_{s-1}}\right]\right\} \cA_{n-1}(1,2,\ldots\hat{s}\ldots, n) + \cO(\delta)
\ee
where $J_i$ denotes the angular momentum operator acting on particle $i$ and $f_s^{\mu\nu} = k_s^\mu a_s^\nu - k_s^\nu a_s^\mu$, as by now usual. 

We would like to extend the analysis of \cite{Bianchi:2014gla} based on the OPE to open string amplitudes with massive states. 

The leading and subleading soft behaviours are captured by the OPE of the soft gluon integrated vertex with the adjacent (integrated) vertices. 
Using 
\be
\int^{z_{s+1}} dz_s (z_{s+1} -z_s)^{2\ap k_s k_{s+1} - 1}  F(z_s, z_i) \approx  {F(z_{s+1}, z_{i}) \over 2\ap k_s k_{s+ 1}} + \ldots \ee
and similarly for $z_{s-1}$ one gets 
\be
V_A(a_s, k_s) V_A(a_{s\pm 1}, k_{s\pm 1})  \approx  \pm {a_s k_{s\pm 1} \over 2 k_s k_{s\pm 1}} V_A(a_{s\pm 1}, k_s + k_{s\pm 1}) + ...
\ee
where $...$ includes massive string states which do not contribute to the leading singularity since
\be
V_A(a_s, k_s) V_A(a_{s\pm 1}, k_{s\pm 1})  \approx  ... + \sum_{M\neq 0} {1 \over 2 k_s k_{s\pm 1} + M^2_H} V_M(H [a_s,a_{s\pm 1}, k_s, k_{s\pm 1}], k_s + k_{s\pm 1}) + ...
\ee
where $V_M$ denotes the vertex operator of a massive state, with momentum $p = k_s + k_{s\pm 1}$ and polarisation $H$ that can be expressed in terms of $a_s,a_{s\pm 1}, k_s, k_{s\pm 1}$.

Expanding the denominator as 
\be 
{1 \over 2 k_s k_{s\pm 1} + M^2_H} \approx {1 \over M^2_H} \left( 1 - {2 k_s k_{s\pm 1}\over M^2_H} + \ldots \right) 
\ee 
one immediately sees that at most the sub-leading (regular $\delta^0$ behaviour) might be affected. However the tri-linear coupling $A{-}A{-}H$ contains at least one soft momentum $k_s$ and this produces a further suppression by $\delta^{{+}1}$. This holds true also for the tachyon since the $T{-}A{-}A$ coupling involves two momenta $\cA_{T{-}A{-}A} = T (k_1k_2 a_1a_2 - k_1a_2 k_2a_1)$, similarly for $H_{\mu\nu}$ since $\cA_{H-A-A} = H_{\mu}{}^\nu f^1_{\nu}{}^\rho f^2_{\rho}{}^\mu + \ldots $, while for $C_{\mu\nu\rho}$ at the first massive level of the superstring  one has 
$\cA_{C-A-A} = C_{\mu\nu\rho} a^\mu_1 a^\nu_2 (k_1 - k_2)^\rho$. 

Let us now consider the case where the soft gluon is attached to a massive (or tachyonic) leg \be
V_A(a_s, k_s) V_M(H_{s\pm 1}, p_{s\pm 1})  \approx {1 \over 2 k_s p_{s\pm 1}} V_{M'}(H' [a_s,H_{s\pm 1}, k_s, p_{s\pm 1}], k_s + p_{s\pm 1}) + ...
\ee
where $M'$ denotes any state at the same mass level as the state $M$. 
For the bosonic string at the tachyonic and first massive level only one kind of particles appears so much so that (for totally symmetric tensors of the first Regge trajectory at level $N=\ell -1$)
\begin{align}
&\cA_{{A}H_\ell H_\ell} = \nonumber\\
&a_1 p_{23} H_2^{\mu_1\ldots\mu_\ell} H_{3,\mu_1\ldots\mu_\ell} + a_{1,\mu}  H_2^{\mu\mu_2\ldots\mu_\ell} p^\nu_{12}H_{3,\nu\mu_2\ldots\mu_\ell} + p_{31,\mu} H_2^{\mu\mu_2\ldots\mu_\ell} a^\nu_{1}H_{3,\nu\mu_2\ldots\mu_\ell} 
+ \cO(\ap p^2) ]
\end{align}
The first term is the string analogue of minimal coupling that is leading in the soft limit $k_1\rightarrow 0$. Gauge invariance fixes the sub-leading term to be as expected. Indeed, for color-ordered amplitudes with $n+1$ gluons and no massive states one finds \cite{Bianchi:2014gla, Bern:2014vva}  
\be
\cA_{n+1}(1,\ldots s \ldots,n+1) = {\pm 1\over 2k_s{\cdot}k_{s\pm1}}  
\cA_{n}(1,\ldots \hat{s}, \ \ldots,n+1)\vert^{a'_{s+1} = a_{s+1} a_s{\cdot}k_{s\pm1}{-}a_s k_s{\cdot}a_{s\pm1}{+}k_s a_s{\cdot}a_{s\pm1}}_{k'_{s+1}=k_{s+1}{+}k_s} + \ldots \ee
expanding in $k_s$ yields
\begin{align}
\pm \left\{ {a_s{\cdot}k_{s\pm1} \over 2k_s{\cdot}k_{s\pm1}} - 
{k_s{\cdot}a_{s\pm1} \over 2k_s{\cdot}k_{s\pm1}} a_s {\cdot} {\partial \over \partial a_{s\pm1}}
 + {a_s{\cdot}k_{s\pm1} \over 2k_s{\cdot}k_{s\pm1}} k_s {\cdot} {\partial \over \partial k_{s\pm1}} + {a_s{\cdot}a_{s\pm1} \over 2k_s{\cdot}k_{s\pm1}} k_s {\cdot} {\partial \over \partial a_{s\pm1}}\right\}\nonumber\\
  \cA_{n}(1,\ldots \hat{s} \ldots,n+1) + \ldots 
 \end{align}
gauge invariance dictates the presence of the additional sub-leading term \be
\mp {k_s{\cdot}k_{s\pm1} \over 2k_s{\cdot}k_{s\pm1}} a_s {\cdot} {\partial \over \partial k_{s\pm1}} \cA_{n}(1,\ldots \hat{s} \ldots,n+1) 
 \ee
that completes at sub-leading order the action of $f_s{:}J_{s{\pm}1}$ on $\cA_{n}(1,\ldots \hat{s} \ldots,n+1)$.

Including $m$ massive states, if the soft gluon is adjacent to two hard gluons the above analysis continues to apply. When at least one of the adjacent legs is massive, let's say the one in position $s+1$, with spin $\ell$ one has 
\be
\cA_{n+1,m}(1,\ldots s \ldots,n+m+1) = {\pm 1\over 2k_s{\cdot}p_{s{+}1}} \cA_{n,m}(1,\ldots \hat{s} \ldots,n+m+1) \vert^{H'_{s+1} = H_{s+1} a_s{\cdot}p_{s{+}1}{+}\ldots}_{p'_{s+1}=p_{s+1}{+}k_s} + \ldots
\ee
where $\ldots$ denotes the additional terms in the tri-linear $V$-$H$-$H$ coupling. Barring a couple of subtleties, we will deal with later on, expanding in $k_s$ one gets (schematically)
\begin{align}
\pm \left\{ {a_s{\cdot}p_{s{+}1} \over 2k_s{\cdot}p_{s{+}1}} - 
\ell {k_s{\cdot}H^{\ldots}_{s{+}1} \over 2k_s{\cdot}p_{s{+}1}} a_s {\cdot} {\partial \over \partial H^{\ldots}_{s{+}1}}
 + {a_s{\cdot}p_{s{+}1} \over 2k_s{\cdot}p_{s{+}1}} k_s {\cdot} {\partial \over \partial p_{s{+}1}} + \ell {a_s{\cdot}H^{\ldots}_{s{+}1} \over 2k_s{\cdot}p_{s{+}1}} k_s {\cdot} {\partial \over \partial H^{\ldots}_{s{+}1}}\right\}\nonumber\\
  \cA_{n,m}(1,\ldots \hat{s} \ldots,n+m+1) + \ldots 
 \end{align}
gauge invariance wrt the soft gluon dictates the presence of the additional sub-leading term \be
\mp {k_s{\cdot}p_{s{+}1} \over 2k_s{\cdot}p_{s{+}1}} a_s {\cdot} {\partial \over \partial p_{s{+}1}} \cA_{n,m}(1,\ldots \hat{s} \ldots,n+m+1) 
 \ee
that completes the action of $f_s{:}J_{s{+}1}$ on $\cA_{n,m}(1,\ldots \hat{s} \ldots,n+m+1)$ at sub-leading order.

Now let us deal with two subtleties: the higher derivative terms in the tri-linear coupling $A$-$H$-$H$ and the possible non-diagonal couplings $A$-$H$-$H'$ that would spoil universality. The former is easy to dispose of, higher derivative corrections to minimal coupling can only affect the sub-leading term that is fixed by gauge invariance wrt the soft gluon starting from the low-derivative terms coded in the OPE.  The latter requires more attention. For open superstrings, as we have seen, already at the first massive level one finds two kinds of particles in the Neveu-Schwarz sector: 
$C_{\mu\nu\rho}$ (3-index anti-symmetric tensor, 84 d.o.f.) and $H_{\mu\nu}$ (2-index symmetric traceless tensor, 44 d.o.f.).
In addition to the `diagonal' couplings $V${-}$C${-}$C$ and $V${-}$H${-}$H$ (and SUSY related) one should consider the mixed coupling $V${-}$H${-}$C$ $\approx \ap M p_{31}{\cdot}H_2{\cdot}C_3{:}[a_1p_{12}]$ that exposes the singular soft factor $1/kp$ since $M_C = M_H$ but gets suppressed by an extra power of the soft momentum in the numerator. Lacking the leading $\delta^{-1}$ term that fixes also the sub-leading $\delta^{0}$ term, thanks to gauge invariance, this kind of higher derivative non-diagonal couplings can at most affect the sub-sub-leading $\delta^{+1}$ (and higher) terms which are not expected to be universal. Although the situation gets exponentially more intricate the higher the mass level and spin, we conclude that no correction are to be expected wrt the standard YM case in the soft behaviour for open string amplitudes involving massive states.  

For illustrative purposes, we will explicitly check the above statements in the soft limit of some 4-point amplitudes with massive string states at the first level. Differently to the case of amplitudes with only mass-less external states that factorise on 3-point amplitudes, that would vanish for real momenta due to collinearity, when some of the external states are massive, the soft limit can produce physical 3-point amplitudes \eg widths or decay rates of massive states into lower mass ones.  

\subsection{Soft limit of $\Amp({A}_1,{T}_2,{T}_3,{T}_4)$}
In this case the limit $k_1\to 0$ is straightforward.
Consider first the expansion of the factor
\begin{equation}
\frac{\Gamma(1+2\alp k_1p_2)\Gamma(1+2\alp k_1p_4)}{\Gamma(1-2\alp k_1p_3)}=\frac{(1+2\alp k_1p_2\psi(1))(1+2\alp k_1p_4\psi(1))}{1-2\alp k_1p_3\psi(1)}=1+\mathcal{O}(\delta^2).
\end{equation}
The expansion of the full amplitude reads
\begin{equation}
\Amp({A}_1,{T}_2,{T}_3,{T}_4)\propto \left(\frac{a_1p_2}{k_1p_2}-\frac{a_1p_4}{k_1p_4}\right) \Amp({T}_2,{T}_3,{T}_4)+ \mathcal{O}(\delta),
\end{equation}
showing the expected singular behavior in both the $s$ and the $t$ channels, whereas the term of order $\mathcal{O}(\delta^0)$ vanishes because 
\begin{equation}
\frac{1}{2}f_1^{\mu\nu}J_{i\mu\nu}\Amp({T}_2,{T}_3,{T}_4)=0,\,\,\,\,i=2,4.
\end{equation}
being the 3-tachyon amplitude a constant independent from the momenta. 

\subsection{Soft limit of $\Amp({A}_1,{A}_2,{T}_3,{T}_4)$}

Consider the amplitude in Eq.~(\ref{AV1V2T3T4}) once we have expressed the Euler beta function in terms of Euler gamma functions

\begin{align}
&\frac{\Gamma(1+2\alp k_1p_4)\Gamma(1+2\alp k_1k_2)}{\Gamma(1-2\alp k_1p_3)}\frac{1}{1-2\alp k_1k_2}\biggr(\frac{k_1p_3}{k_1k_2}(-a_1a_2+2\alp(a_1p_3\,\,a_2p_3+a_1p_4\,\,a_2p_4)\nonumber\\
&-a_1p_3\,\,a_2p_4\frac{1+2\alp k_1p_4}{k_1k_2}-a_1p_4\,\,a_2p_3\frac{1+2\alp k_1p_3}{k_1p_4\,\,k_1k_2}k_1p_3\biggr).
\end{align}
As already verified, the combination of Euler gamma functions in the above expression contributes in the limit $\delta\to 0$ as $1+\mathcal{O}(\delta^2)$, thus can be
neglected.
We have two terms of order ${\delta^{-1}}$:
\begin{equation}
-\frac{a_1p_3\,\,a_2p_4}{k_1k_2}-\frac{a_1p_4\,\,a_2p_3\,\,k_1p_3}{k_1k_1\,\,k_1p_4}.
\label{O(1/delta)}
\end{equation}
Using the identity
\begin{equation}
\frac{k_1p_3}{k_1k_2\,\,k_1p_4}=-\frac{1}{k_1k_2}-\frac{1}{k_1p_4},
\end{equation}
Eq.~(\ref{O(1/delta)}) can be written as
\begin{align}
&\frac{a_1p_4\,\,a_2p_3 - a_1p_3\,\,a_2p_4}{k_1k_2}+\frac{a_1p_4\,\,a_2p_3}{k_1p_4}=\biggr(\frac{1}{2}a_1p_+a_2p_- - \frac{1}{2}a_1p_-a_2p_+\biggr)\frac{1}{k_1k_2}\nonumber\\
&+\biggr(a_1\frac{p_+-p_-}{2}\,\,a_2\frac{p_-}{2}+ a_1\frac{p_+-p_-}{2}\,\,a_2\frac{p_+}{2}\biggr)\frac{1}{k_1p_4}\nonumber\\
&=\biggr(-\frac{a_1k_2}{k_1k_2}+\frac{a_1p_4}{k_1p_4}\biggr)a_2\frac{p_{34}}{2} +\frac{1}{2}\frac{a_1p_-\,\,a_2k_1}{k_1k_2}-\frac{1}{2}\frac{a_1p_4\,\,a_2k_1}{k_1p_4},
\end{align}
where 
\begin{equation}
p_+=p_3+p_4\,\,\,\,\,\,\,\,p_-=p_3-p_4.
\end{equation}
The leading soft contribution is, as expected, 
\begin{equation}
\biggr(-\frac{a_1k_2}{k_1k_2}+\frac{a_1p_4}{k_1p_4}\biggr)\Amp({A}_2,{T}_3,{T}_4).
\end{equation}
The order $\mathcal{O}(\delta^0)$ contribution to the amplitude reads
\begin{align}
&-a_1a_2\frac{k_1p_3}{k_1k_2} +\frac{1}{2}\frac{a_1p_-\,\,a_2k_1}{k_1k_2}-\frac{1}{2}\frac{a_1p_4\,\,a_2k_1}{k_1p_4}
 +2\alp\biggr(\frac{k_1p_3}{k_1k_2}(a_1p_3\,\,a_2p_3+a_1p_4\,\,a_2p_4+a_1p_4\,\,a_2p_3)\nonumber\\
 &+a_1p_4\,\,a_2p_3\biggr(\frac{k_1p_3}{k_1p_4} + \frac{k_1k_2}{k_1p_4} + 1\biggr)-a_1p_3\,\,a_2p_4\biggr(\frac{k_1p_4}{k_1k_2} +1\biggr) \biggr)\\
 &=-a_1a_2\frac{k_1p_3}{k_1k_2} +\frac{1}{2}\frac{a_1p_-\,\,a_2k_1}{k_1k_2}-\frac{1}{2}\frac{a_1p_4\,\,a_2k_1}{k_1p_4}+2\alp a_1p_+a_2p_+\\
&=-\frac{1}{2}a_1a_2\frac{k_1p_3}{k_1k_2} +  \frac{1}{2}a_1a_2\frac{k_1p_4}{k_1k_2} + \frac{1}{2}a_1a_2\frac{k_1p_4}{k_1p_4} +\frac{1}{2}\frac{a_1p_-\,\,a_2k_1}{k_1k_2}-\frac{1}{2}\frac{a_1p_4\,\,a_2k_1}{k_1p_4}+\mathcal{O}(\delta).
\end{align}
In the above expression we recognize the expected behavior
\begin{align}
&\frac{1}{2k_1k_2}f_{1\mu\nu}J_2^{\mu\nu}\Amp(V_2,{T}_3,{T}_4)=\biggr(a_1a_2\,\,k_1\frac{p_{34}}{2}-a_1\frac{p_{34}}{2}\,\,a_2k_1\biggr)\frac{1}{k_1k_2},\\
&\frac{1}{2k_1p_4}f_{1\mu\nu}J_4^{\mu\nu}\Amp({A}_2,{T}_3,{T}_4)=\biggr(\frac{1}{2}a_1a_2\,\,k_1p_4-a_1p_4\,\,a_2k_1\biggr)\frac{1}{k_1p_4}.
\end{align}

\subsection{Soft limit of $\Amp({T}_1,{T}_2,A_3,{H}_4)$}

Consider the amplitude in Eq.~(\ref{eq:TTVH}). Let us first discuss the soft limit of the kinematical factor
\begin{align}
&\frac{\Gamma(-1+2\alp k_3p_4)\Gamma(1+2\alp p_3k_3)}{\Gamma(-2\alp p_1k_3)}=\frac{\Gamma(1+2\alp k_3p_4)\Gamma(1+2\alp p_3k_3)}{\Gamma(1-2\alp p_1k_3)}\frac{k_3p_1}{k_3p_4(1-2\alp k_3p_4)}\nonumber\\
&=(1+\cO(\delta^2))\frac{k_3p_1}{k_3p_4}\frac{1}{1-2\alp k_3p_4}.
\end{align}
It is then convenient to multiply the tensorial part of the amplitude for the above expansion in order to identify more easily the contributions up to $\mathcal{O}(\delta^0)$:
\be
\frac{a_3p_4}{k_3p_4}p_2{H}p_2=\frac{a_3p_4}{k_3p_4}\frac{p_{12}}{2}{H}\frac{p_{12}}{2} +\cO(\delta^0), \quad  \cO(\delta^0)=\frac{a_3p_4}{k_3p_4}k_3{H}\frac{p_{12}}{2};
\ee
\be
-\frac{a_3p_2}{k_3p_2}p_2{H}p_2=-\frac{a_3p_2}{k_3p_2}\frac{p_{12}}{2}{H}\frac{p_{12}}{2} +\cO(\delta^0), \quad  \cO(\delta^0)=-\frac{a_3p_2}{k_3p_2}k_3{H}\frac{p_{12}}{2};
\ee
\be
2a_3p_2\,\,p_2{H}k_3\frac{p_1k_3}{p_2k_3\,\,k_3p_4}=-2\frac{a_3p_2}{p_2k_3}p_2{H}k_3-2\frac{a_3p_2}{k_3p_4}p_2{H}k_3;
\ee
\be
-2a_3{H}p_2\frac{p_1k_3}{k_3p_4}\frac{p_2k_3}{p_2k_3}=2a_3{H}p_2+2\frac{k_3p_2}{k_3p_4}a_3{H}p_2.
\ee
The leading order $\cO(\delta^{-1})$ behaves as expected
\be
\cA^{\delta^{-1}}({T}_1,{T}_2,A_3,{H}_4) =\biggr(\frac{a_3p_4}{k_3p_4}-\frac{a_3p_2}{k_3p_2}\biggr)\,\, \frac{p_{12}}{2}{H}\frac{p_{12}}{2},
\ee
being $\Amp({T}_1,{T}_2,H_4)= \frac{p_{12}}{2}{H}\frac{p_{12}}{2}$.
Look at the subleading contribution:
\begin{align}
&\frac{1}{k_3p_2}\biggr(-a_3p_2\,\,k_3{H}\frac{p_{12}}{2}-2a_3p_2\,\,p_3{H}k_3+a_3{H}p_2\,\,k_3p_2\biggr)\nonumber\\
&\frac{1}{k_3p_4}\biggr(a_3p_4\,\,k_3{H}\frac{p_{12}}{2}-2a_3p_2\,\,p_3{H}k_3+a_3{H}p_2(2k_3p_2+k_3p_4)\biggr)\nonumber\\
&=\frac{1}{k_3p_2}\biggr(a_3p_2\,\,k_3{H}\frac{p_{12}}{2} - k_3p_2\,\,a_3{H}\frac{p_{12}}{2}\biggr)-\frac{1}{k_3p_4}\biggr(2a_3\frac{p_{12}}{2}\,\,k_3{H}\frac{p_{12}}{2}-2k_3\frac{p_{12}}{2}\,\,a_3{H}\frac{p_{12}}{2}\biggr).
\end{align}

It is easy to verify that the above expressions coincide with:
\be
\frac{1}{2k_3p_i}f_{3\mu\nu}J^{\mu\nu}_i\,\,\Amp({T}_1,{T}_2,H_4),\quad i=2,4.
\ee
We recall that the angular momentum for a spin-2 particle takes the form
\be
J_{\mu\nu}=p_\mu \frac{\de}{\de p^\nu}-p_\nu\frac{\de}{\de p^\mu}+2H_{\mu\lambda}\frac{\de}{\de H^\nu_\lambda}-2H_{\nu\lambda}\frac{\de}{\de H^\mu_\lambda}.
\ee

\subsection{Soft limit of $\cA({A}_1,{A}_2,{A}_3,C_4)$}

The amplitude is given in Eq~(\ref{eq:AAAC}).
Let us study the soft behaviour when $k_3 \rightarrow 0$. Recall that $s+t+u = 1/\ap$

The Veneziano factor $\cB(1,1)$ yields
\be
\cB(1,1) = { \Gamma (1-\ap s)  \Gamma (1-\ap t)  \over  \Gamma (1 + \ap u) } = { \Gamma (2 \ap k_3 p_4)  \Gamma (1+ 2\ap k_3 k_2)  \over  \Gamma (1- 2\ap k_3 k_1) } \approx  {1 \over 2 \ap k_3 p_4} \times [1 + \cO(\delta^2)]
\ee
To leading order the polarisation dependent factor yields
\be
\cP = - a_1{\cdot}C_4{:}f_2 \left(a_3{\cdot}k_1 -  {k_3{\cdot}k_1 \over k_3{\cdot}k_2} a_3{\cdot}k_2 \right)
\ee
combining the two one gets, as expected 
\be
\cA^{\delta^{-1}}_4({A}_1,{A}_2,{A}_3,C_4) =  \left({a_3{\cdot}p_4 \over k_3 p_4} - {a_3{\cdot}k_2 \over k_3{\cdot}k_2}  \right) a_1{\cdot}C_4{:}f_2 
= \left({a_3{\cdot}p_4 \over k_3 p_4} - {a_3{\cdot}k_2 \over k_3{\cdot}k_2}  \right) \cA_3({A}_1,{A}_2,C_4)
\ee
To sub-leading order $\delta^0$, one gets
\be
\cA^{\delta^{0}}_4({A}_1,{A}_2,{A}_3,C_4) = \left({2 a_1{\cdot}C_4{\cdot}f_3{\cdot}f_2 - a_1{\cdot}f_3{\cdot} C_4{\cdot}f_2\over k_3 p_4} - {2 a_1{\cdot}C_4{\cdot}f_3{\cdot}f_2 \over k_3{\cdot}k_2}  \right)  \nonumber \ee \be = \left({f_3{:}J_4 \over k_3 p_4} - {f_3{:}J_2 \over k_3{\cdot}k_2}  \right) \cA_3({A}_1,{A}_2,C_4)
\ee
as expected, where $${J_2}^{\mu}{}_{\nu} = {k_2^\mu  \partial \over \partial k_2^\nu} -  {k_2^\mu \partial \over \partial  k_2^\nu} + a_2^\mu {\partial \over \partial a_2^\nu} - a_2^\mu {\partial \over \partial a_2^\nu} \: , \quad {J_4}^{\mu}{}_{\nu} = p_4^\mu {\partial \over \partial p_4^\nu} - p_4^\mu {\partial \over \partial  p_4^\nu} + 3 C_4^{\mu\lambda\rho} {\partial \over \partial C_4^{\nu\lambda\rho}} - 3 C_4^{\nu\lambda\rho} {\partial \over \partial C_4^{\mu\lambda\rho}} $$ 
Actually ${\partial/\partial p_4}$ acts trivially in this case]. 

With little more effort one would get the same result for $k_1\rightarrow 0$, while for $k_2\rightarrow 0$ the only contributions come from `standard' soft behaviour of gluons hitting adjacent gluons.

This gives support to our general conclusion that superstring amplitudes with $n$ massless and $m$ massive external legs on the boundary of the disk behave universally in the soft limit. 

\subsection{Soft limit of $\cA({A}_1,{A}_2,{A}_3,{H}_4)$}

The amplitude is given in Eq~(\ref{eq:AAAH}).
Let us study the soft behaviour when $k_3 \rightarrow 0$. Recall that $s+t+u = 1/\ap$ as for  $\cA({A}_1,{A}_2,{A}_3,C_4)$. Following the same steps one finds to leading order
\be
\cA^{\delta^{-1}}_4({A}_1,{A}_2,{A}_3,H_4) =   {k_3k_1 \over k_3p_4} 
\left({a_3k_2 \over k_3k_2} - {a_3k_1 \over k_3k_1}  \right) 
\left[{1\over 2} m_H^2 a_1Ha_2 + a_2k_1\, a_1Hk_2 - a_1f_2Hk_2\right]  =  \nonumber
\ee
\be
- \left({a_3k_2 \over k_3k_2} - {a_3p_4 \over k_3p_4}  \right) f_1Hf_2 + \cO(1) = 
\left({a_3k_2 \over k_3k_2} - {a_3p_4 \over k_3p_4}  \right) \cA_3({A}_1,{A}_2,H_4)
\ee
where use of $m_H^2 = -p_4^2 = -2k_1k_2 + \cO(\delta)$ has been made.

At subleading order one finds several terms \ie

\begin{align}
&\frac{1}{k_3k_2} \biggr[ a_2a_3\left( k_3p_4\,a_1Hk_2-a_1Hk_3\frac{m^2_H}{2}\right) + a_1Ha_3a_2k_3\frac{m^2_H}{2}+a_1Hk_2\,a_2k_3\,a_3k_1\nonumber\\
&-a_1Hk_3\,a_2k_1\,a_3k_2-a_1f_2f_3Hk_2+a_1f_3f_2k_2-k_3p_4\,a_1Ha_2\,a_3k_2+a_3p_4\,a_2Ha_2\,k_2k_3\nonumber\\
&+a_2a_3\,a_1Hk_2\,k_2k_3+a_2k_1\,a_1Ha_3\,k_2k_3\biggr]\nonumber\\
+&\frac{1}{k_3p_4}[ a_2a_3\,a_1Hk_3\frac{m^2_H}{2}+a_1Ha_3\,a_2k_3\frac{m^2_H}{2}+a_1Ha_3\,a_2k_1\,k_3k_2-a_1f_2Ha_3\,k_3k_1-a_1f_3Ha_2\nonumber\\
&-a_1Hk_3\,a_2k_1\,a_3k_2  -a_1f_2Hk_3\,a_3k_2 - a_1f_3Hk_2\,a_2k_1-a_1f_2f_3Hk_2-a_3p_4\,a_1f_2Hk_3]
\end{align}

Summing the terms within squared parenthesis everything can be written in terms of $f_1$, $f_2$, $f_3$ (as dictated by gauge invariance at this order) finding
\be
\cA^{\delta^{0}}_4({A}_1,{A}_2,{A}_3,H_4) =  \left({f_1 H f_3f_2 - f_1 H f_2f_3 \over k_3k_2} - {f_1 f_3 H f_2 - f_1 H f_3f_2 \over k_3p_4}  \right) \nonumber \ee \be = 
\left({f_3J_2 \over k_3k_2} - {f_3J_4 \over k_3p_4}  \right) 
\cA_3({A}_1,{A}_2,H_4)
\ee
with ${J_2}^\mu{}_\nu$ given above and 
$$
{J^H_4}^{\mu}{}_{\nu} = p_4^\mu {\partial \over \partial p_4^\nu} - p_4^\mu {\partial \over \partial  p_4^\nu} + 2 H_4^{\mu\lambda} {\partial \over \partial H_4^{\nu\lambda}} - 2 H_4^{\nu\lambda} {\partial \over \partial H_4^{\mu\lambda}} $$
As above ${\partial /\partial p_4}$ acts trivially in this case.
This gives further support to our general arguments on the soft limit.

\section{Holomorphic soft limit}
  
In this section we verify that  open string amplitudes with massive external states enjoy the same universal behaviour as YM amplitudes in the holomorphic soft limit \cite{He:2014bga}. In this limit the holomorphic spinor $u_s$ of a positive helicity gluon (inserted between leg 1 and leg $n$) is scaled to zero $u=\delta\hat{u}_s$. In SYM the leading behaviour as $\delta^{-2}$ is governed by the operator
$$
S^0_{YM} = {\lef n1 \re \over \lef ns \re \lef s1 \re}
$$
the sub-leading behaviour as $\delta^{-1}$ is governd by the operator
$$
S^1_{YM} = {\lef n1 \re \over \lef ns \re \lef s1 \re} \left\{  {\lef s n \re \over \lef 1n \re} \bar{u}^{\dot\alpha}_s {\partial\over \partial \bar{u}^{\dot\alpha}_1}  + {\lef s 1 \re \over \lef n1 \re} \bar{u}^{\dot\alpha}_s {\partial\over \partial \bar{u}^{\dot\alpha}_n}\right\} 
$$
For MHV amplitude the sub-leading term vanishes and the procedure exponentiates \cite{He:2014bga}. In general it is convenient to use momentum conservation to express two $\bar{u}$'s in terms of the remaining ones and the ${u}$'s. In our case, an obvious choice is to express $\bar{u}_4$ and $\bar{u}_5$ that appear in the definition of the massive momentum 
$p = k_4 + k_5 = u_4\bar{u}_4+ u_5\bar{u}_5$. When taking derivatives one has to take into account the mass constraint 
$m^2 = - (k_4 + k_5)^2$ as we will see momentarily.

\subsection{$\cA(A^+,A^+,A^+,{C})$}
\label{AAACsec}

Consider the amplitude 
\be
\cA(1^+,2^+,3^+,{C}_0) = B(1-\alp s,1-\alp t)\frac{[13]m_{C}}{\lef 12 \re \lef 23 \re}
\label{ApApApC}
\ee
and take the limit for ${u}_2\to \delta {u}_2$, with $\delta\to 0$. It is straightforward to show that 
\be
B(1-\alp s,1-\alp t)=\frac{1}{\alp \lef 13 \re [13]}(1+\cO(\delta^2)).
\ee
The momentum of the massive particle is the sum of two massless momenta $p=k_4+k_5$ with the constraint $\lef 45 \re [54]=m_{C}^2$. 
This constraint implies
\begin{align}
m_{C}^2=	\lef 45 \re [54]=\lef 13 \re [13] + \delta(\lef 12 \re [12]+ \lef 23 \re [23]),\\
m_{C}=m_{C}(\delta=0)\left(1+\frac{1}{2}\delta\left(\frac{\lef 12 \re [12]}{\lef 13 \re[13]}+\frac{\lef 23 \re [23]}{\lef 13 \re [13]}\right) \right)+\cO(\delta^2).
\end{align}
Expanding the amplitude, one finds
\begin{align}
&\cA(1^+,2^+,3^+,{C}_0)=\frac{1}{\delta^2}\frac{m_{C}^2}{\lef 31 \re [13]}\frac{[13]m_{C}}{\lef 12 \re \lef 23 \re}(1+\cO(\delta^2))\nonumber\\
&=-\left(1+\delta(\lef 12 \re [12]+\lef 23 \re [23])\right)\frac{[13]m_{C}}{\lef 12 \re \lef 23 \re}\left( 1+\frac{\delta}{2 \lef 13\re [13]}(\lef 12 \re [12]+\lef 23 \re[23]) \right)+\cO(\delta^2)\nonumber\\
&=-\frac{1}{\delta^2}\frac{[13]m_{C}}{\lef 12 \re \lef 23 \re}\left(1+\frac{3}{2}\delta\left(\frac{\lef 12 \re [12]}{ \lef 13\re [13]}+\frac{\lef 23 \re[23]}{ \lef 13\re [13]}\right)\right).
\end{align}
The leading contribution to the holomorphic soft limit is easily identified to be
\be
\cA^{(-2)}(1^+,2^+,3^+,{C}_0)=\frac{1}{\delta^2}\frac{\lef 31\re}{\lef 32 \re \lef 21 \re}\cA(1^+,3^+,{C}_0)=\frac{1}{\delta^2}\frac{\lef 31\re}{\lef 32 \re \lef 21 \re}\frac{[13]^2}{m_{C}}=-\frac{[13]m_{C}}{\delta^2\lef 12 \re \lef 23 \re}.
\ee
that meets our expectations. 

The sub-leading contribution is expected to be
\be
\cA^{(-1)}(1^+,2^+,3^+,{C}_0)=\frac{1}{\delta}\frac{\lef 31\re}{\lef 32 \re \lef 21 \re}\left( \frac{\lef 23\re}{\lef 13\re}{\bar{u}}_2\frac{\de}{\de{\bar{u}}_1}+ \frac{\lef 21\re}{\lef 31\re}{\bar{u}}_2\frac{\de}{\de{\bar{u}}_3}\right)\cA(1^+,3^+,{C}_0).
\ee
In the presence of the mass constraint, the derivatives wrt ${\bar{u}}_{1, 3}$ are replaced by
\be
\frac{\de}{\de{\bar{u}}_{1,3}}\to \frac{d}{d{\bar{u}}_{1,3}} = \frac{\de}{\de{\bar{u}}_{1,3}}+\frac{\de m_{C}}{\de {\bar{u}}_{1,3}}\frac{\de}{\de m_{C}},
\ee
with
\be
\frac{\de m_{C}}{\de {\bar{u}}_1}=\frac{\lef 13 \re}{2m_{C}}|3],\quad\frac{\de m_{C}}{\de {\bar{u}}_3}=-\frac{\lef 13 \re}{2m_{C}}|1].
\ee
Writing the three-point function $\cA(1^+,3^+,{C}_0)$
in a slightly different way
\be
\frac{[13]^2}{m_{C}}=\frac{m_{C}^3}{\lef 13\re^2}+\cO(\delta),
\ee
we need to evaluate only the derivative of the tri-linear coupling respect to $m_{C}$. Finally we find
\begin{align}
&\cA(1^+,2^+,3^+,{C}_0)^{(-1)}=\frac{1}{\delta}\left( \frac{\lef 13\re [23]}{2\lef 21 \re m_{C}} - \frac{\lef 13 \re [21]}{2\lef 32 \re m_{C}}\right)\frac{\de}{\de m_{C}}\cA(1^+,3^+,{C}_0)\nonumber\\
&=-\frac{1}{\delta}\frac{3}{2}\frac{[13]m_{C}}{\lef 12 \re \lef 23 \re}\left(\frac{\lef 23 \re[23]}{\lef 13 \re [13]}+\frac{\lef 12 \re[12]}{\lef 13 \re [13]}\right).
\end{align}
that exposes the expected behaviour, too.

In order to complete our analysis, we consider the case in which the soft momentum is $k_3$. 
Let's first expand the amplitude in Eq.~\eqref{ApApApC} up to the order  $\delta^{-1}$
\be
\cA(1^+,2^+,3^+,{C}_0)=\frac{1}{\delta^2}\frac{[13]m_0^3}{2k_3p_4\,\lef 12 \re \lef 23 \re}\left(1+\frac{3}{2}\delta \frac{\lef 13\re [13]+\lef 23 \re [23]}{\lef 12 \re [12]}\right)+\cO(\delta^0).
\ee
At leading order, the soft operator is simply
\be
\frac{a_3^+k_2}{2k_3k_2}-\frac{a_3^+p_4}{k_3p_4},
\ee
which in the spinor helicity formalism becomes
\be
\frac{\lef q2\re }{\lef 32 \re\lef 3q\re}-\frac{\lef q4 \re[43]+\lef q5\re [53]}{2k_3p_4\lef 3q\re}=\frac{\lef 12 \re [13]}{\lef 23 \re \,2k_3p_4},
\ee
with the help of Schouten's identity.
The expected leading order behavior looks like
\be
\cA^{(-2)}(1^+,2^+,3^+,{C}_0)=\frac{1}{\delta^2}\frac{\lef 12 \re [13]}{\lef 23 \re \,2k_3p_4}\cA(1^+,2^+,C_0)=\frac{1}{\delta^2}\frac{[13]m_0^3}{\lef 12 \re \lef 23 \re\,2k_3p_4},
\label{ApApApCexpected}
\ee
where we exploited the fact that
\be
\cA(1^+,2^+,C_0)=\frac{[12]^2}{m_0}.
\ee
At sub-leading order we expect the soft operator to be
\be
\frac{f_3^+{:}J_2}{2k_2k_3}-\frac{f_3^+{:}J_4}{2k_3p_4}\rightarrow \frac{1}{\lef 23\re}\bar u_3 \frac{d}{d \bar u_2}-\frac{1}{2k_3p_4}\left([34]\bar u_3 \frac{d}{d\bar u_4}+[35]\bar u_3\frac{d}{d\bar u_5}\right).
\ee
Noticing that
\begin{align}
&\left(\frac{\de}{\de \bar u_2}+\frac{\de m_0}{\de \bar u_2} \frac{\de}{\de m_0} \right) \frac{[12]^2}{m_0}=\frac{3}{2}\frac{[12]\bar u_1}{m_0}\nonumber\\
&\left(\frac{\de}{\de \bar u_4}+\frac{\de m_0}{\de \bar u_4} \frac{\de}{\de m_0} \right) \frac{[12]^2}{m_0}=\frac{\lef 45 \re [12]^2 \bar u_5}{2m_0^3}\nonumber\\
&\left(\frac{\de}{\de \bar u_5}+\frac{\de m_0}{\de \bar u_5} \frac{\de}{\de m_0} \right) \frac{[12]^2}{m_0}=-\frac{\lef 45 \re [12]^2 \bar u_4}{2m_0^3}\nonumber,
\end{align}
we find
\be
\cA^{(-1)}(1^+,2^+,3^+,{C}_0)=\frac{3}{2\delta}\frac{[12][13]}{\lef 23 \re m_0},
\ee
which is compatible with Eq.~\eqref{ApApApCexpected} after noticing that the sub-leading term in the expansion can be written as
\be
\frac{3}{2\delta}\frac{[13][23][12]}{2k_3p_4}+\frac{3}{2\delta}\frac{[13][23][12]}{m_0}\left(-\frac{1}{2k_3p_4}+\frac{1}{\lef 23 \re[23]}\right)=\frac{3}{2\delta}\frac{[12][13]}{\lef 23 \re m_0}.
\ee

\subsection{$\cA(A^-,A^+,A^-,{C})$}

Consider now the amplitude 
\be
\cA(1^-,2^+,3^-,{C}_0)=\frac{\Gamma(1+2\alp k_1k_2)\Gamma(1+2\alp k_2k_3)}{\Gamma(1-2\alp k_2p_4)}\frac{\lef 13 \re^3}{\lef 12 \re \lef 23 \re m_{C}}.
\ee
Taking the limit in which ${u}_2\to 0$, we have
\be
\cA(1^-,2^+,3^-,{C}_0)=\frac{1}{\delta^2}\frac{\lef 13 \re^3}{\lef 12 \re \lef 23 \re m_{C}(\delta=0)}\left( 1-\frac{\delta}{2}\left( \frac{\lef 12 \re [12]}{\lef 13 \re[13]}+\frac{\lef 23 \re [23]}{\lef 13 \re [13]} \right) \right).
\ee
For the leading term one finds
\be
\cA^{(-2)}(1^-,2^+,3^-,{C}_0)=\frac{1}{\delta^2}\frac{\lef 13 \re}{\lef 12 \re \lef 23 \re}\frac{\lef 13 \re^2}{m_{C}},
\ee
for the sub-leading term 
\be
\cA^{(-1)}(1^-,2^+,3^-,{C}_0)=\frac{1}{\delta}\frac{\lef 13 \re}{\lef 12 \re \lef 23 \re}\frac{\lef 13 \re^2}{m_{C}}\left(\frac{\lef 23 \re \lef 13 \re}{\lef 13 \re 2m_{C}}[23] - \frac{\lef 21 \re \lef 13 \re}{\lef 31 \re 2m_{C}} [21]\right)\frac{\lef 13 \re^2}{m^2_{C}}.
\ee
that behaves as expected in the holomorphic soft limit.

In this case we will not consider the limit in which $k_3\to 0$ since the three-point amplitude $\cA(1^-,2^+,C_0)=0$ vanishes.

\subsection{$\cA(A^-,A^+,A^+,{H}^{++})$}

Consider finally the amplitude
\be
\cA(1^-,2^+,3^+,{H}^{++})=\cB(1-\alp s,1-\alp t)\frac{\lef 14 \re^4 [13]}{m_{H} \lef 12 \re \lef 23 \re \lef 45 \re^2}.
\label{AmApApH}
\ee
Expanding for ${u}_2\to 0$, one finds
\be \cA(1^-,2^+,3^+,{H}^{++})=\frac{1}{\delta^2}\frac{\lef 14 \re^4}{\lef 12 \re \lef 23 \re \lef 31 \re \lef 45 \re^2}m_{H}(\delta) =\frac{1}{\delta^2}\frac{\lef 14 \re^4 [45]^2}{\lef 12 \re \lef 23 \re \lef 31 \re m_{H}^3(\delta)}.
\ee
Using 
\be
\lef 14 \re [45]= \lef 13 \re [35]+\delta \lef 12 \re [25],
\ee
we have
\be
\cA(1^-,2^+,3^+,{H}^{++})=\frac{1}{\delta^2}\frac{\lef 14 \re^2[35]^2 \lef 31 \re}{\lef 12 \re \lef 23 \re m_{H}(\delta)^3}\left( 1+ 2\delta \frac{\lef 12 \re[25]}{\lef 13 \re [35]}-\frac{3}{2}\delta\left( \frac{\lef 12 \re [12]}{\lef 13 \re[13]}+\frac{\lef 23 \re [23]}{\lef 13 \re [13]}  \right)\right).
\label{eq:SoftAAAH++}
\ee
At this stage the soft limit appears straightforward.
The leading term reads
\be
\cA^{(-2)}(1^-,2^+,3^+,{H}^{++})=\frac{\lef 31 \re}{\lef 12 \re \lef 23 \re}\frac{\lef 14 \re^2 [35]^2}{m_{H}^3}
\ee
Using the expressions for the two derivatives
\begin{align}
\frac{d}{d{\bar{u}}_1}\cA_3(1^-,3^+,{H}^{++})&=-\frac{3}{2 m_{H}^5}\lef 14\re^2[35]^2\lef 13 \re {\bar{u}}_3\\
\frac{d}{d{\bar{u}}_3}\cA_3(1^-,3^+,{H}^{++})&=\frac{2}{m_{H}^3}\lef 14 \re^2 [35]{\bar{u}}_5+\frac{3}{2 m_{H}^5}\lef 14\re^2[35]^2\lef 13 \re {\bar{u}}_1
\end{align}
into the soft sub-leading term
\be
\cA^{(-1)}(1^-,2^+,3^+,{H}^{++})=\frac{\lef 31 \re}{\lef 12 \re \lef 23 \re}\left(\frac{\lef 23 \re}{\lef 13 \re}{\bar{u}}_2\frac{d}{d{\bar{u}}_1} +\frac{\lef 21 \re}{\lef 31 \re} {\bar{u}}_2\frac{d}{d{\bar{u}}_3}\right)\frac{\lef 14 \re^2 [35]^2}{m_{H}^3}
\ee
we reproduce exactly Eq.~(\ref{eq:SoftAAAH++}).

Let's consider the limit in which $k_3\to 0$. Expanding the amplitude in Eq.~\eqref{AmApApH} up to the order $\delta^{-1}$ we find
\be
\cA(1^-,2^+,3^+,{H}^{++})=\frac{1}{\delta^2}\frac{\lef 14 \re^2 \lef 12 \re[13] [25]^2}{2k_3p_4\,\lef 23 \re m_0^3}\left(1-\frac{3}{2}\delta \frac{\lef 13 \re [13]+\lef 23 \re [23]}{\lef 12 \re [12]} +2\delta \frac{\lef 13 \re[35]}{\lef 12 \re [25]}\right).
\ee
Using the leading order soft operator we derived in Sec.~\ref{AAACsec}, we find that
\be
\cA^{(-2)}(1^-,2^+,3^+,{H}^{++})=\frac{1}{\delta^2}\frac{\lef 12 \re [13]}{\lef 23 \re \,2k_3p_4}\cA(1^-,2^+,{H}^{++})=\frac{1}{\delta^2}\frac{\lef 14 \re^2 \lef 12 \re[13] [25]^2}{2k_3p_4\,\lef 23 \re m_0^3},
\ee
using
\be
\cA(1^-,2^+,{H}^{++})=\frac{\lef 14\re^2[25]^2}{m_0^3}.
\ee
The sub-leading soft behavior of the amplitude is determined by
\begin{align}
&\frac{[32]}{2k_2k_3}\bar u_3\frac{d}{d\bar u_2}\frac{\lef 14 \re^2[25]^2}{m_0^3}=\frac{3}{2\delta}\frac{\lef 14 \re^2[13][25]^2}{\lef 23 \re m_0^3},\nonumber\\
&\frac{1}{2k_3p_4} \left( [34]\bar u_3 \frac{d}{d\bar u_4}+[35] \bar u_3 \frac{d}{d \bar u_5} \right) \frac{\lef 14 \re^2[25]^2}{m_0^3}=2\frac{\lef 14\re^2 [23][35]}{2k_3p_4 m_0^3}.
\end{align}
Following the same algebraic manipulations as in Sec.~\ref{AAACsec} it can be shown that these two terms reproduce the sub-leading soft term of the expansion in Eq.~\eqref{AmApApH}.

\section{Conclusions}

We have computed several open bosonic and super- string scattering amplitudes on the disk with massive and tachyonic external states in critical dimension as well as in $D=4$ (for the superstring, using the spinor helicity basis). 

We have then checked their universal behaviour when massless gluons go soft, despite the presence of higher derivative couplings, and offered a general argument to this effect based on world-sheet OPE. We have also checked consistency of the factorisation on the first massive pole of the MSS formula obtained in \cite{Mafra:2011nv, Mafra:2011nw} relating open superstring amplitudes on the disk to tree-level SYM amplitudes. 

We have only briefly considered closed strings. For gravitons, even in the presence of massive external legs, one would expect a universal soft behaviour up to sub-sub-leading order ($\delta^{+1}$) \cite{Weinberg:1965nx, Gross:1968in, Cachazo:2014fwa, Schwab:2014sla, Zlotnikov:2014sva, Kalousios:2014uva}
\be
\cM_{n}(1,2,\ldots,s,\ldots,n) =  \ee
\be \sum_{i\neq s} \left[{k_i{\cdot}h_s{\cdot}k_{i} \over k_s{\cdot}k_{i}} + {2 k_i {\cdot} h_s{\cdot}J_{i}k_s \over k_s{\cdot}k_{i}} + {k_s{\cdot}J_i{\cdot}h_s{\cdot}J_{i}{\cdot}k_s \over k_s{\cdot}k_{i}}\right] \cM_{n-1}(1,2,\ldots\hat{s}\ldots, n) + \cO(\delta^2)
\ee
This should hold true at tree-level and with the understanding that interactions be governed by minimal couplings. 
While in closed Type II superstrings on the sphere the soft limit of amplitudes with massless states is the same as in gravity at tree level, for bosonic  strings -- and in fact for the heterotic string, too -- 
the presence of a $\phi R^2$ vertex with the dilaton spoils the universal behave even at leading order, in that a soft graviton attached to a hard graviton can produce a hard dilaton thus producing a mixed amplitude\footnote{As suggested in \cite{DiVecchia:2015oba}, one may be tempted to propose a generalisation of the soft theorem whereby dilatons and gravitons are `unified' into a gravi-dilaton with symmetric transverse but non-traceless polarisation tensor $e_{\mu\nu} = h_{\mu\nu} {+} \phi_{\mu\nu}$ with $\phi_{\mu\nu} = \eta_{\mu\nu} {-} k_\mu\bar{k}_\nu {-} k_\nu\bar{k}_\mu$ and $\bar{k}^2 = 0$ $\bar{k}k=1$. Yet for the Kalb-Ramond anti-symmetric tensor $b_{\mu\nu}$ which is odd under world-sheet parity $\Omega$, one expects a vanishing behaviour at leading order \cite{Bianchi:2014gla}.}

Using KLT relations \cite{Kawai:1985xq} one can efficiently compute closed amplitudes with massive external states as `squares' of open string amplitudes with massive external states, like the ones we have considered in the present investigation. We plan to carry out this analysis in simple cases and study the soft behaviour at tree level confirming universality, respectively lack of it, in the case of the closed superstring (both Type IIA and Type IIB),  respectively in the case of the bosonic or heterotic string due to the presence of the $\phi R^2$ terms \cite{MBAG2}. We hope to shed further light on the soft behaviour of the Kalb-Ramond field, the dilaton \cite{Ademollo:1975pf} and the other moduli fields \cite{Chen:2014cuc}. It would also be interesting to investigate the soft behaviour of loop amplitudes and to test the validity of the new proposal \cite{Stieberger:2014cea, Stieberger:2015qja} of getting the graviton from the collinear limit of two gluons beyond tree level and in the presence of massive external states.

 \acknowledgments
We would like to thank P.~Di Vecchia, C.~Mafra, R.~Marotta, M.~Mojaza, O.~Schlotterer, S.~Stieberger, C-K~Wen for discussions.

\appendix
\section{Open bosonic string 3-point amplitudes}
\label{subsec:3bosonic}

For the sake of completeness we summarize all the possible three point functions involving open bosonic string states up to the first massive level, Eqs.~\ref{eq:tac}, \ref{eq:vec}, \ref{eq:H}.
Kinematics of three point on-shell amplitudes is fixed in terms of the masses of the particles involved in the process.
This property will be used repeatedly and stressed wherever necessary.
In the following formulas a factor $(2\pi)^D\delta^{D}\left(\sum_i p_i\right)$, with $D\leq 26$, resulting from integration over the zero mode of the coordinate fields $X^\mu$, is always understood. We will also drop a factor of $g_s\,(\alp)^{(D/2-3)/2}$, which is $g_s\,(\alp)^5$ for the bosonic string in critical dimension, 
 but, following the discussion in section~\ref{sec:ChanPaton}, we will explicitly include the relevant Chan-Paton factors $f_{abc}$ or $d_{abc}$ that make the full `amplitude' Bose symmetric.

\medskip

$\bullet$ {\bf $TTT$ vertex}
\begin{align}
\mathcal{A}(T_1,T_2,T_3)&{=}d_{abc} \left\langle c\,e^{ip_1X}(z_1)\,\,c\,e^{ip_2X}(z_2)\,\,c\,e^{ip_3X}(z_3)\right\rangle\nonumber\\
&{=}d_{abc} \,\,z_{12}z_{13}z_{23}\,\,z_{12}^{2\alp p_1p_2}z_{13}^{2\alp p_1p_3}z_{23}^{2\alp p_2p_3}{=}d_{abc},
\end{align}
where we used the identity $(p_i{+}p_j)^2{=}{-}2m^2_T {+} 2p_ip_j{=}{-}m^2_T$, so that $2\alp p_ip_j{=}m_T^2{=}{-}\frac{1}{\alp}$ for all $i$, $j$. 
The symbol $z_{ij}$ stands for $z_i {-} z_j$.
In order to simplify the notation, from now on we will introduce
the notation
\begin{equation}
P_i^\mu{=}\sum_{j\neq i}\frac{p_j^\mu}{z_{ji}}.
\label{eq:P}
\end{equation}
In general, $P_i$ is contracted always with the $i${-}th polarization vector/tensor. 
Exploiting `transversality' \ie $p_i^\mu t^i_{\mu \ldots} {=}0$, we will always replace the sum in Eq.~\eqref{eq:P} with:
\be
P_1{=}\frac{p_{23}}{2}\frac{z_{23}}{z_{12}z_{13}},\hspace{5mm}P_2{=}\frac{p_{31}}{2}\frac{z_{13}}{z_{12}z_{23}},\hspace{5mm}P_3{=}\frac{p_{12}}{2}\frac{z_{12}}{z_{13}z_{23}}.
\ee
$\bullet$ {\bf $TT A$ vertex}
\begin{align}
\Amp({A}_1,T_2,T_3)&{=} \frac{1}{\sqrt{2\alp}}f_{abc} \left\langle c\,a_{1\mu}\,i\de X^\mu\,e^{ik_1X}(z_1)\,\,c\,e^{ip_2X}(z_2)\,\,c\,e^{ip_3X}(z_3)\right\rangle\,\,\nonumber\\
&{=}\sqrt{2\alp}\, f_{abc} \,\, z_{12}z_{13}z_{23}\,\,a_{1\mu}P_1^\mu\,\,\produtt{=}\sqrt{2\alp}\,f_{abc} \,\, \frac{1}{2}a_1p_{23}.
\end{align}
$\bullet$ {\bf $T A A$ vertex}
\begin{align}
\Amp({A}_1,{A}_2,T_3)&{=} \frac{1}{2\alp}d_{abc} \left\langle c\,a_1i\de X\,e^{ik_1X}(z_1)\,\,c\,a_2i\de X\, e^{ik_2X}(z_2)\,\,c\,e^{ip_3X}(z_3)\right\rangle\nonumber\\
&{=} d_{abc} \,\,\left(2\alp\, a_1\frac{p_{23}}{2}\,\,a_2\frac{p_{31}}{2} {+} \,a_1a_2\right).
\end{align}
The amplitude can be rewritten in a manifestly gauge invariant form:
\be
\Amp({A}_1,{A}_2,T_3){=} 2\alp\,d_{abc} \, \left( a_1\frac{p_{23}}{2}\,\,a_2\frac{p_{31}}{2} {+} a_1a_2\,\,k_1k_2\right){=} 2\alp\,d_{abc}\,\frac{1}{2} f_{1\mu\nu}f_2^{\nu\mu}.
\ee
$\bullet$ {\bf $AAA$ vertex}
\begin{align}
&\Amp( {A}_1, {A}_2, {A}_3)
{=}\frac{1}{(2\alp)^{3/2}} f_{abc} \left\langle c\,a_1i\de X\,e^{ik_1X}(z_1)\,\,c\,a_2i\de X\,e^{ik_2X}(z_2) \,\,c\,a_3i\de X\,e^{ik_3 X}(z_3) \right\rangle\nonumber\\
&{=}\sqrt{2\alp} f_{abc} \left( a_1a_2\,\,a_3\frac{k_{12}}{2} {+} a_1a_3\,\,a_2\frac{k_{31}}{2}{+}a_2a_3\,\,a_1\frac{k_{23}}{2}{+}2\alp a_1\frac{k_{23}}{2}\,\,a_2\frac{k_{31}}{2}\,\,a_3\frac{k_{12}}{2} \right).
\end{align}
$\bullet$ {\bf  $T A H$ vertex}
\begin{align}
\Amp({A}_1,T_2,\cH_3)&{=} \frac{1}{(2\alp)^{3/2}}\,f_{abc} \left\langle c\,a_1i\de X\,e^{ik_1X}(z_1)\,\,c\,e^{ip_2X}(z_2)\,\,c\,i\de X\,H_3\, i\de X\, e^{ip_3X}(z_3)\right\rangle\nonumber\\
&{=}\sqrt{2\alp}\,f_{abc}\,\, \left(2a_1H_3\frac{p_{12}}{2} {+} 2\alp a_1\frac{p_{23}}{2}\,\,\frac{p_{12}}{2}H_3\frac{p_{12}}{2}\right).
\end{align}
$\bullet$ {\bf $AAH$ vertex}
\begin{align}
&\Amp({A}_1,{A}_2,H_3)
{=}\frac{d_{abc}}{(2\alp)^2} \left\langle c\,a_1i\de X e^{ik_1X}(z_1)\,\,c\,a_2i\de X e^{ik_2X}(z_2)\,\,c\,i\de XH_3 i\de X e^{ip_3 X}(z_3)\right\rangle\nonumber\\
&{=}d_{abc}\biggr(2a_1H_3a_2{ {+}} 2\alp\biggr(2a_1\frac{p_{23}}{2}\,\,a_2H_3\frac{p_{12}}{2}{{+}}2a_2\frac{p_{31}}{2}\,\,a_1H_3\frac{p_{12}}{2}{{+}}a_1a_2\,\,\frac{p_{12}}{2}H_3\frac{p_{12}}{2}\biggr) \nonumber\\
&{{-}}(2\alp)^2  a_1\frac{p_{23}}{2} \,\,a_2\frac{p_{13}}{2}\,\,\frac{p_{12}}{2}H_3\frac{p_{12}}{2} \biggr).
\end{align}
One can rewrite the above amplitude in the manifestly gauge invariant form
\be
\cA({A}_1,{A}_2,H_3) {=} 2\alp\,d_{abc}\,\, \biggr(2\,\tr(f_1H_3f_2){-}\alp \tr(f_1f_2)\,\,k_1H_3k_2\biggr).
\label{eq:V1V2H3}
\ee
$\bullet$ {\bf $AHH$ vertex}
\begin{align}
\Amp({A}_1,H_2,H_3)
&{{=}}\frac{f_{abc}}{(2\alp)^{5/2}}\left\langle c a_1i\de Xe^{ik_1X}(z_1)\,\,c\,i\de XH_2i\de Xe^{ip_2X}(z_2)\,\,c\,i\de XH_3i\de Xe^{ip_3X}(z_3)\right\rangle\nonumber\\
&{{=}}\sqrt{2\alp}\,f_{abc}\,\,\biggr(2a_1\frac{p_{23}}{2}\,\,\tr(H_2H_3){{-}}4\tr(f_1H_2H_3){+}4\alp\frac{p_{31}}{2}H_2f_1H_3\frac{p_{12}}{2}\nonumber\\
&{{+}}8\alp a_1\frac{p_{23}}{2}\,\,\frac{p_{31}}{2}H_2H_3\frac{p_{12}}{2}
{{+}}(2\alp)^2a_1\frac{p_{23}}{2}\,\,\frac{p_{31}}{2}H_2\frac{p_{31}}{2}\,\,\frac{p_{12}}{2}H_3\frac{p_{12}}{2}\biggr).
\end{align}
$\bullet$ {\bf $THH$ vertex}
\begin{align}
\Amp(T_1,H_2,H_3)&{=} \frac{1}{(2\alp)^2}d_{abc} \left\langle c\,e^{ip_1X}(z_1)\,\,c\,i\de X\,H_2 \,i\de X e^{ip_2X}(z_2)\,\,c\,i\de X\,H_3\, i\de X e^{ip_3X}(z_3)\right\rangle\nonumber\\
&{=}d_{abc}\,\,\left(2\,\tr(H_2H_3){+}4(2\alp)\frac{p_{31}}{2}H_2H_3\frac{p_{12}}{2}{+}(2\alp)^2 \frac{p_{31}}{2}H_2\frac{p_{31}}{2}\,\frac{p_{12}}{2}H_3\frac{p_{12}}{2}\right).
\end{align}
$\bullet$ {\bf $TTH$ vertex}
\begin{align}
&\Amp(T_1,T_2,H_3){=}\frac{1}{2\alp} d_{abc} \left\langle c\,e^{ip_1X}(z_1)\,\,c\, e^{ip_2X}(z_2)\,\,c\,i\de X \,H_3 \,i\de X e^{ip_3X}(z_3)\right\rangle\nonumber\\
&{=}d_{abc}\,\,z_{12}z_{13}z_{23}\,\,\produtt\,\,2\alp P_3H_3P_3{=}2\alp\,d_{abc}\,\, \frac{p_{12}}{2}H_3\frac{p_{12}}{2}.
\end{align}
$\bullet$ {\bf $HHH$ vertex}
\begin{align}
&\Amp(H_1,H_2,H_3)\nonumber\\
&{=}\frac{d_{abc}}{(2\alp)^3} \left\langle c\,i\de X\,H_1\,i\de X\,e^{ip_1X}(z_1)\,\,c\,i\de X\,H_2\,i\de X\,e^{ip_2X}(z_2)\,\,c\,i\de X\,H_3\,i\de X\,e^{ip_3X}(z_3)\right\rangle\nonumber\\
&{=}d_{abc}\,\,\biggr(8\,\tr(H_1H_2H_3){+}2\alp\biggr(\tr(H_1H_2)\,\,\frac{p_{12}}{2}H_3\frac{p_{12}}{2}{+}\tr(H_1H_3)\,\,\frac{p_{31}}{2}H_2\frac{p_{31}}{2}\nonumber\\
&{+}\tr(H_2H_3)\,\,\frac{p_{23}}{2}H_1\frac{p_{23}}{2}{+}8\frac{p_{23}}{2}H_1H_2H_3\frac{p_{12}}{2}{+}8\frac{p_{23}}{2}H_1H_3H_2\frac{p_{31}}{2}{+}8\frac{p_{31}}{2}H_2H_1H_3\frac{p_{12}}{2}\biggr)\nonumber\\
&{+}(2\alp)^2\biggr(\frac{p_{23}}{2}H_1H_2\frac{p_{31}}{2}\,\,\frac{p_{12}}{2}H_3\frac{p_{12}}{2} 
{+}\frac{p_{23}}{2}H_1H_3\frac{p_{12}}{2}\,\,\frac{p_{31}}{2}H_2\frac{p_{31}}{2}{+}\frac{p_{31}}{2}H_1H_3\frac{p_{12}}{2}\,\,\frac{p_{23}}{2}H_1\frac{p_{23}}{2}\biggr)\nonumber\\
&{+}(2\alp)^3\frac{p_{23}}{2}H_1\frac{p_{23}}{2}\,\,\frac{p_{31}}{2}H_2\frac{p_{31}}{2}\,\,\frac{p_{12}}{2}H_3\frac{p_{12}}{2}\biggr).
\end{align}

\section{Open bosonic string four{-}point amplitudes}

In this appendix we sketch the computation of the open bosonic string amplitudes involving massive and tachyonic states. For simplicity we consider color{-}ordered amplitudes. Complete amplitudes arise after multiplying by the relevant Chan{-}Paton factors $\tr(t^1t^2t^3t^4)$ and summing over non{-}cyclic permutations. In fact, exploting `twist symmetry' \ie $\Omega$ invariance, one can reduce the sum to three terms (instead of six). 
Exploiting conformal invariance we choose to fix $z_1\to \infty$, $z_2{=}1$, $z{=}\frac{z_{12}z_{34}}{z_{13}z_{24}}$ and $z_4{=}0$.
A factor $g^2_{s,ap}(\alp)^{(D/2{-}4)/2}\,(2\pi)^D\,\delta^D(\sum_i p_i)$ is always understood.

\medskip
$\bullet$ {\bf Veneziano amplitude ($TTTT$)}
\begin{align}
&\Amp(T_1,T_2,T_3,T_4){=}\left\langle c\,e^{ip_1X}(z_1)\,\,c\,e^{ip_2X}(z_2)\,\,\int dz_3\,e^{ip_3X}(z_3)\,\,c\,e^{ip_4X}(z_4)\right\rangle\nonumber\\
&{=}z_{12}z_{14}z_{24}\int dz_3\,\produtt{=}\int_0^1 dz\,(1{-}z)^{2\alp p_2p_3}z^{2\alp p_3p_4}{=}B(1{{+}}2\alp p_2p_3,1{{+}}2\alp p_3p_4).
\label{eq:TTTT}
\end{align}
Introducing the Mandelstam variables $(p_1{+}p_2)^2{=}(p_3{+}p_4)^2{=}{-}s$, $(p_2{+}p_3)^2{=}(p_1{+}p_4)^2{=}{-}t$, $(p_1{+}p_3)^2{=}(p_2{+}p_4)^2{=}{-}u$,  
we can rewrite the Veneziano amplitude as
\begin{equation}
\Amp(T_1,T_2,T_3,T_4){=}\frac{\Gamma({-}1{-}\alp s)\Gamma({-}1{-}\alp t)}{\Gamma({-}2{-}\alp (s{+}t))}.
\end{equation}
$\bullet$ {\bf $ATTT$ amplitude}
\begin{align}
\Amp({A}_1,T_2,T_3,T_4)&{=}\frac{1}{\sqrt{2\alp}}\left\langle c\,a_{1}i\de X e^{ik_1 X}(z_1)\,\, c\,e^{ip_2X}(z_2)\,\,\int dz_3\,e^{ip_3X}(z_3)\,\,c\,e^{ip_4X}(z_4)\right\rangle\nonumber\\
&{=}\frac{1}{\sqrt{2\alp}}\left(\frac{a_1p_2}{ k_1p_2} {-} \frac{a_1p_4}{ k_1p_4} \right)\frac{\Gamma(1{+}2\alp k_1p_2)\Gamma(1{+}2\alp k_1p_4)}{\Gamma(1{-}2\alp k_1p_3)}.
\end{align}
$\bullet$ {\bf $AATT$ amplitude}
\begin{align}
&\Amp({{A}_1},{{A}_2},T_3,T_4){=}\frac{1}{2\alp}\left\langle ca_{1}i\de X e^{ik_1X}(z_1)\,\,ca_{2}\,i\de X e^{ik_2 X}(z_2)\int dz_3e^{ip_3X}(z_3)\,\,ce^{ip_4X}(z_4)\right\rangle\nonumber\\
&=\biggr(a_1a_2{-}2\alp(a_1p_3\,\,a_2p_3 {+} a_1p_4\,\,a_2p_4){+}a_1p_3\,\,a_2p_4\frac{1{+}2\alp k_1p_4}{k_1p_3}{+}a_1p_4\,\,a_2p_3\frac{1{+}2\alp k_1p_3}{k_1p_4}\biggr)\nonumber\\
&B(1{+}2\alp k_1p_4,{-}1{+}2\alp k_1k_2).
\label{AV1V2T3T4}
\end{align}
$\bullet$ {\bf $AAAT$ amplitude}
\begin{align}
&\Amp({A}_1,{A}_2,T_3,{A}_4){=}\int dz_3 \left\langle c a_2 i\de X e^{ip_1 X} (z_1) \,\, c a_1i\de X\, e^{ip_2 X} (z_2) \,\,e^{ip_3 X} (z_3) \,\, c a_4 i\de  X e^{ip_4 x} (z_4)\right\rangle \nonumber\\
&{=}\frac{\Gamma(1{+}2\alp p_1 p_2)\Gamma(1{+}2\alp p_1 p_4)}{\Gamma(1{-}2\alp p_1 p_3)} \frac{p_1 p_3}{2\alp p_1 p_2 \,\, p_1 p_4} \biggr\{ {-} a_1 a_2 \,\, a_4 p_1 \frac{1{+}2\alp p_1 p_3}{2\alp p_3  p_4} {+}a_1 a_2 \,\, a_4  p_2 \frac{1{+}2\alp p_2 p_3}{2\alp p_3  p_4}\nonumber\\
& {+} a_1 a_4 \,\, a_2 p_1 \frac{1{+}2\alp p_1 p_3}{2\alp p_2 p_3} {-}  a_1 a_4 \,\, a_2 p_4 \frac{1{+}2\alp p_3 p_4}{2\alp p_2 p_3} {-} a_2 a_4 \,\, a_1 p_2 \frac{1{+}2\alp p_2 p_3}{2\alp p_1 p_3} {+}  a_2 a_4 \,\, a_1 p_4 \frac{1{+}2\alp p_3 p_4}{2\alp p_1 p_3} \nonumber\\
& {+}2\alp\biggr( a_1 p_4 \,a_2 p_1 \,\, a_4 p_2 {-} a_1 p_2 \,a_2 p_4 \,\, a_4 p_1 {+} a_1 p_2 \,a_2 p_1 \,a_4 p_1 \frac{1{+}2\alp p_1 p_3}{2 \alp p_3 p_4} {-}a_1 p_4 \,a_2 p_1 \,a_4 p_1 \frac{1{+}2\alp p_1 p_3}{2\alp p_2 p_3} \nonumber\\
& {-} a_1 p_4 \,\, a_2 p_4 \,\, a_4 p_2  \frac{1{+}2\alp p_3 p_4}{2\alp p_1 p_3} {+} a_1 p_2 \,\, a_2 p_4 \,\, a_4 p_2 \frac{1{+}2\alp p_2 p_3}{2\alp p_1 p_3} {+} a_1 p_4 \,\, a_2 p_4 \,\, a_4 p_1 \frac{1{+}2\alp p_3 p_4}{2\alp p_2 p_3}  \nonumber\\
& {-} a_1 p_2 \,\, a_2 p_1 \,\, a_4 p_2 \frac{1{+}2\alp p_2 p_3}{2\alp p_3 p_4} \biggr) \biggr\}.
\end{align}
$\bullet$ {\bf $TT A H$ amplitude}
\begin{align}
&\Amp(T_1,T_2,{A}_3,H_4){=}\frac{1}{(2\alp)^{\frac{3}{2}}}\biggr\langle ce^{ip_1X}(z_1)\,ce^{ip_2X}(z_2)\int dz_3\, a\,i\de X e^{ikX}(z_3)\,ci\de XHi\de Xe^{ip_4X}(z_4)\biggr\rangle\nonumber\\
&{=}\biggr({-}2a_3Hp_2{-}2a_3Hk_3\frac{1{+}2\alp k_3p_1}{2{-}2\alp k_3p_4}{+}2\alp a_3p_4\biggr(p_2Hp_2\frac{1{-}2\alp k_3p_4}{2\alp k_3p_1}{+}k_3Hk_3\frac{1{+}2\alp k_3p_1}{2{-}2\alp k_3p_4}{+}2p_2Hp_3\biggr)\nonumber\\
&{-}2\alp a_3p_2\biggr(\frac{k_3p_4\,\,p_2Hp_2}{2\alp p_2k_3\,p_1k_3}(1{-}2\alp k_3p_4){-}p_3Hp_3\frac{1{+}2\alp p_1k_3}{2\alp p_2k_3}{-}2p_2Hk_3\frac{1{-}2\alp k_3p_4}{2\alp p_2k_3}\biggr)\biggr)\nonumber\\
&\frac{\Gamma({-}1{+}2\alp k_3p_4)\Gamma(1{+}2\alp p_2k_3)}{\Gamma({-}2\alp p_1k_3)}.
\label{eq:TTVH}
\end{align}

\section{Open superstring 3{-}point amplitudes}
\label{app:supervertex}

In this section we compute all the possible tri{-}linear couplings involving superstring states up to the first massive level in the Neveu{-}Schwarz sector following the same conventions as in Appendix~\ref{subsec:3bosonic}.

$\bullet$ {\bf $AAC$ vertex}
\begin{align}
&\Amp({A}^{(0)}_1,{A}^{({-}1)}_2,C^{({-}1)}_3)\nonumber\\
&{=}\frac{d_{abc}}{\sqrt{2\alp}}\left\langle c\left(a_1\,i\de X {+} k_1\psi\,\,a_1\psi\right)\,e^{ik_1X}(z_1)\,\,ce^{{-}\phi}a_2\psi \,e^{ik_2X}(z_2)\,\,ce^{{-}\phi}C_3\psi\psi\psi\, e^{ip_3X}(z_3)\right\rangle\nonumber\\
&{=}\sqrt{2\alp}d_{abc} \,\,\frac{z_{12}z_{13}z_{23}}{z_{23}}\,\,\produtt\,\,6\frac{a_2^\mu C_{3\mu\nu\rho}a_1^\nu k_1^\rho}{z_{13}^2z_{23}}
{=}6\sqrt{2\alp}\,d_{abc} \,\,a_2^\mu  C_{3\mu\nu\rho}  a_1^\nu  \frac{k_{12}^\rho}{2}.
\label{eq:VVC}
\end{align}
$\bullet$ {\bf $AAH$ vertex}
\begin{align}
&\Amp({A}^{(0)}_1,{A}^{({-}1)}_2,H^{({-}1)}_3)\nonumber\\
&{=}\frac{d_{abc}}{2\alp}\langle c\left(a_1\,i\de X {+} k_1\psi\,\,a_1\psi\right)\,e^{ik_1X}(z_1)\,\,ce^{{-}\phi}\,a_2\psi e^{ik_2X}(z_2)\,\,ce^{{-}\phi}\, i\de X\,H_3\,\psi\,e^{ip_3X}(z_3)\rangle\nonumber\\
&{=}{-}2\alp d_{abc}\,\tr(f_1H_3f_2).
\end{align} 
$\bullet$ {\bf $AH C$ vertex}
\begin{align}
&\Amp({A}^{({-}1)}_1,H^{(0)}_2,C^{(0)}_3)\nonumber\\
&{=}\frac{f_{abc}}{2\alp}\left\langle c(a_1i\de X{+}k_1\psi\,a_1\psi)e^{ik_1X}(z_1)\,\,ce^{{-}\phi}i\de X H_2\psi e^{ip_2X}(z_2)\,\,ce^{{-}\phi}C_3\psi\psi\psi e^{ip_3X}(z_3)\right\rangle\nonumber\\
&{=}2\alp f_{abc} \,\,\frac{z_{12}z_{13}z_{23}}{z_{23}}\,\,\produtt\,\,6P_2^\rho\,\,H_{2\rho\sigma}C_{3\mu\nu\lambda} \frac{\eta^{\mu\sigma}a_1^\nu k_1^\lambda}{z_{23}z_{13}^2}{=}f_{abc} 12\alp \,\frac{p_{31}}{2}H_2C_3a_1\frac{p_{12}}{2}.
\end{align}
$\bullet$ {\bf $AHH$ vertex}
\begin{align}
&\Amp({A}^{(0)}_1,H^{({-}1)}_2,H^{({-}1)}_3)\nonumber\\
&{=}\frac{f_{abc}}{(2\alp)^{3/2}}\left\langle c(a_1i\de X{+}k_1\psi\,a_1\psi)e^{ik_1X}(z_1)\,\,ce^{{-}\phi}i\de X\,H_2\psi e^{ip_2X}(z_2)\,\,ce^{{-}\phi}i\de X\,H_3\psi\,e^{ip_3X}(z_3) \right\rangle \nonumber\\
&{=}\sqrt{2\alp} f_{abc}
\biggr(2\tr(f_1H_2H_3) {+} a_1\frac{p_{23}}{2}\tr(H_2H_3){+}2\alp \biggr(\frac{p_{31}}{2}H_2f_1H_3\frac{p_{12}}{2} {+} a_1\frac{p_{23}}{2}\,\frac{p_{31}}{2}H_2H_3\frac{p_{12}}{2}\biggr)\biggr).
\end{align}
$\bullet$ {\bf $ACC$ vertex}
\begin{align}
&\Amp({A}_1^{(0)},C_2^{({-}1)},C^{({-}1)}_3)\nonumber\\
&{=}\frac{f_{abc}}{\sqrt{2\alp}}\left\langle c(a_1i\de X{+}k_1\psi\,a_1\psi)e^{ik_1X}(z_1)\,\,ce^{{-}\phi}C_2\psi\psi\psi e^{ip_2X}(z_2)\,\,ce^{{-}\phi}C_3\psi\psi\psi\,e^{ip_3X}(z_3) \right\rangle\nonumber\\
&{=}\sqrt{2\alp}f_{abc} 6\biggr(a_1\frac{p_{23}}{2} \tr(C_2C_3){-}3\,\tr(f_1C_2C_3)\biggr).
\end{align}
$\bullet$ {\bf $AAA$ vertex}
\begin{align}
&\Amp({A}^{(0)}_1,{A}^{({-}1)}_2,{A}^{({-}1)}_3)\nonumber\\
&{=}\frac{f_{abc}}{\sqrt{2\alp}}\left\langle c (z_1)\,(a_1i\de X{+}k_1\psi\,\,a_1\psi)e^{ik_1X}(z_1)\,\,c\,e^{{-}\varphi}\,a_2\psi\,e^{ik_2X}(z_2)\,c\,e^{{-}\varphi}\,a_3\psi\,e^{ik_3X}(z_3)\right\rangle\nonumber\\
&{=}\sqrt{2\alp}f_{abc}\biggr(a_1\frac{k_{23}}{2}\,\,a_2a_3{+}a_2\frac{k_{31}}{2}\,\,a_1a_3{+}a_3\frac{k_{12}}{2}\,\,a_1a_2\biggr).
\end{align}
$\bullet$ {\bf $HHH$ vertex}
\begin{align}
&\Amp(H^{(0)}_1,H^{({-}1)}_2,H^{({-}1)}_3){=}\frac{d_{abc}}{(2\alp)^2}\langle c\,H_{1\mu\nu} (i\de X^\mu i\de X^\nu {+} p_1\psi\,\psi^\mu i\de X^\nu {+} \de \psi^\mu \psi^\nu)\,e^{ip_1X}(z_1)\nonumber\\
&ce^{{-}\varphi}\,i\de X H_2 \psi\,e^{ip_2X}(z_2)\,\,c\,e^{{-}\varphi}\,i\de X H_3 \psi\,e^{ip_3X}(z_3)  \rangle\nonumber\\
&{=}3\tr(H_1H_2H_3){+}2\alp\biggr(3\frac{p_{23}}{2}H_1H_2H_3\frac{p_{12}}{2}{+}3\frac{p_{23}}{2}H_1H_3H_2\frac{p_{31}}{2}{+}3\frac{p_{31}}{2}H_2H_1H_3\frac{p_{12}}{2}\nonumber\\
&{+}\tr(H_2H_3)\frac{p_{23}}{2}H_1\frac{p_{23}}{2}{+}\tr(H_1H_3)\frac{p_{31}}{2}H_2\frac{p_{31}}{2}{+}\tr(H_1H_2)\frac{p_{12}}{2}H_3\frac{p_{12}}{2}\biggr)\nonumber\\
&{+}(2\alp)^2\biggr(\frac{p_{23}}{2}H_1\frac{p_{23}}{2}\,\,\frac{p_{31}}{2}H_2H_3\frac{p_{12}}{2}{+}\frac{p_{31}}{2}H_2\frac{p_{31}}{2}\,\,\frac{p_{23}}{2}H_1H_3\frac{p_{12}}{2}{+}\frac{p_{12}}{2}H_3\frac{p_{12}}{2}\,\,\frac{p_{23}}{2}H_1H_2\frac{p_{31}}{2}\biggr).
\end{align}
$\bullet$ {\bf $CCC$ vertex}
\begin{align}
&\Amp(C_1^{(0)},C_2^{({-}1)},C_3^{({-}1)})\nonumber\\
&{=}\frac{d_{abc}}{\sqrt{2\alp}}\left\langle  c\, C_{1\mu\nu\rho}(i\de X^\mu {+} p\psi\,\psi^\mu)\psi^\nu\psi^\rho\,e^{ip_1X}(z_1)\,\,c\,e^{{-}\varphi} C_2\psi\psi\psi\,e^{ip_2X}(z_2)\,\,c\,e^{{-}\varphi}\,\psi\psi\psi\,e^{ip_3X}(z_3) \right\rangle\nonumber\\
&{=}\sqrt{2\alp}d_{abc}\biggr(\frac{p_{23}^\mu}{2}C_{1\mu\nu\rho}C_2^{\rho\sigma\lambda}C_{3\lambda\sigma}^{\,\,\,\,\,\,\,\,\,\,\nu} {+} \frac{p_{31}^\mu}{2}C_{2\mu\nu\rho}C_3^{\rho\sigma\lambda}C_{1\lambda\sigma}^{\,\,\,\,\,\,\,\,\,\,\nu} {+}\frac{p_{12}^\mu}{2}C_{3\mu\nu\rho}C_1^{\rho\sigma\lambda}C_{3\lambda\sigma}^{\,\,\,\,\,\,\,\,\,\,\nu}\biggr).
\end{align}
$\bullet$ {\bf $CCH$ vertex}
\begin{align}
&\Amp(C^{(0)}_1,C^{({-}1)}_2,H^{({-}1)}_3)\nonumber\\
&{=}\frac{d_{abc}}{2\alp}\left\langle  c C_{1\mu\nu\rho}(i\de X^\mu {+} p\psi\,\psi^\mu)\psi^\nu\psi^\rho e^{ip_1X}(z_1)\,\,ce^{{-}\varphi}\,C_2\psi\psi\psi e^{ip_2X}(z_2)\,\, ce^{{-}\varphi} i\de X H_3 \psi e^{ip_3X}(z_3) \right\rangle\nonumber\\
&d_{abc}\,\,6\tr(C_1C_2H_3){+}2\alp\biggr( 6\tr(C_1C_2)\frac{p_{12}}{2}H_3\frac{p_{12}}{2} {+} \frac{p_{23}}{2}C_1C_2H_3\frac{p_{12}}{2}{+}6\frac{p_{12}}{2}H_3C_1C_2\frac{p_{31}}{2}  \biggr).
\end{align}
$\bullet$ {\bf $CHH$ vertex}
\begin{align}
&\Amp(C^{(0)}_1,H^{({-}1)}_2,H^{({-}1)}_3)\nonumber\\
&{=}\frac{d_{abc}}{(2\alp)^{3/2}}\left\langle  cC_{1\mu\nu\rho}(i\de X^\mu {+} p\psi\,\psi^\mu)\psi^\nu\psi^\rho e^{ip_1X}(z_1)\,ce^{{-}\varphi}i\de X H_2 \psi\,e^{ip_2X}(z_2)\, ce^{{-}\varphi}i\de X H_3 \psi e^{ip_3X}(z_3) \right\rangle\nonumber\\
&{=}\sqrt{2\alp}d_{abc}\biggr(2\frac{p_{23}^\mu}{2}C_{1\mu\nu\rho}H_{2\sigma}^\nu H_3^{\rho\sigma}{+}2C_{1\mu\nu\rho}H_3^{\mu\nu}H_{2\sigma}^\rho\frac{p_{31}^\sigma}{2}{+}2C_{1\mu\nu\rho}H_{2\mu\rho}H_{3\sigma}^\nu\frac{p_{12}^\sigma}{2}\nonumber\\
&{+}4\alp\frac{p_{23}}{2}C_{1\mu\nu\rho}H_{2\sigma}^\nu H_{3\lambda}^\rho \frac{p_{31}^\sigma}{2}\frac{p_{12}^\lambda}{2}\biggr).
\end{align}

\section{Open superstring four{-}point amplitudes}

Let us discuss the derivation of the 4{-}point amplitudes with one massive external state, \emph{i.e.}
$\cA(A_1,A_2,A_3,H_4)$ and $\cA(A_1,A_2,A_3,C_4)$.
\subsection{$\cA_{AAAC}$ amplitude}
\label{app:AAAC}

With a judicious choice of super{-}ghost pictures and c{-}ghost insertion one has
\be
\cA(A^{({-}1)}_1,A^{(0)}_2,A^{(0)}_3,C^{({-}1)}_4) {=} \lim_{(z_1,z_2,z_4) {\rightarrow} (\infty, 1, 0)} \int_0^1 dz_3 \nonumber
\ee
\be
\langle c e^{{-}\varphi} a_1\psi e^{ik_1X}(z_1) c (a_2 \de X {+} i k_2\psi a_2\psi)e^{ik_2X}(z_2)
(a_3 \de X {+} i k_3\psi a_3\psi)e^{ik_3X}(z_3) c C_4 \psi\psi\psi e^{ip_4X}(z_4)\rangle.
\ee
There are only two kinds of non{-}vanishing contractions:
$\langle \psi(1) {:}\psi\psi{:}(2)  \de X(3) {:}\psi\psi\psi{:}(4)\rangle {+} (2\leftrightarrow 3)$ 
and $\langle \psi(1) {:}\psi\psi{:}(2)  {:}\psi\psi{:}(3) {:}\psi\psi\psi{:}(4)\rangle$. 
The first kind of contractions yields
\be
{1\over 2 z_1} [a_1{\cdot}C_4{:}f_2 a_3{\cdot} P_3(x) {+} {1\over x^2} a_1{\cdot}C_4{:}f_3 a_2{\cdot} P_2(x)], \ee
where setting $z_3 {=} x$ we also have
\be
P_3(x) {=} {-}{1\over x} \left(k_1 {+} {k_2\over 1{-}x}\right) \quad  \quad P_2(x) {=} {x \over 1{-} x} k_3 {-} k_1.\ee
It is convenient to define also
\be
P_4(x) {=} {-}\left(k_2 {+} {1\over x}k_3\right) \quad  \quad P_1(x) {=} k_2 {+} x k_3.\ee
The second kind of contractions yields
\be
{-} {1\over 2 z_1} \left( {a_1{\cdot}f_2{\cdot}C_4{:}f_3 \over x^2}  {+}  {a_1{\cdot}f_3{\cdot}C_4{:}f_2 \over x} {+} 2 {a_1{\cdot}\dot{C}_4\overline{{\cdot}f_3{\cdot}}\dot{f}_2 \over x(1{-}x)} \right)\ee
in a self{-}explanatory index{-}free notation. 
Including the momentum factor and the (super-)ghost correlators yields (barring $\delta(\Sigma p)$ etc) 
\begin{align}
&\cA(A^{({-}1)}_1,A^{(0)}_2,A^{(0)}_3,C^{({-}1)}_4) {=} \int_0^1 dx x^{{-}\ap s {+} 1} (1{-}x)^{{-}\ap t} \left[
{-} a_1{\cdot}C_4{:}f_2\left(a_3{\cdot}k_1 {+} {a_3{\cdot} k_2\over 1{-}x}\right) \right . \nonumber\\
& \left. {+} a_1{\cdot}C_4{:}f_3 \left( {a_2{\cdot}k_3 \over 1{-}x} {-} {a_2{\cdot} k_1\over x}\right) 
{+} {a_1{\cdot}f_2{\cdot}C_4{:}f_3 \over x}  {+}  {a_1{\cdot}f_3{\cdot}C_4{:}f_2} {+} 2 {a_1{\cdot}\dot{C}_4\overline{{\cdot}f_3{\cdot}}\dot{f}_2 \over 1{-}x} \right].
\end{align}
Perusing the factorial properties of $\Gamma(z)$, finally yields 
\begin{align}
\cA_{{A}{A}{A}C} {=}  \cB (1,1) \biggr\{ {-} a_1{\cdot}C_4{:}f_2 \left[ a_3{\cdot}k_1 {-} {u \over t} {a_3{\cdot} k_2} \right]
{-} a_1{\cdot}C_4{:}f_3 \left[ {u \over t} {a_2{\cdot}k_3} {-} {u \over s} {a_2{\cdot} k_1}\right] \nonumber\\
{+} {u \over s} {a_1{\cdot}f_2{\cdot}C_4{:}f_3 }  {-}  {a_1{\cdot}f_3{\cdot}C_4{:}f_2} 
{-} 2 {u \over t} {a_1{\cdot}\dot{C}_4\overline{{\cdot}f_3{\cdot}}\dot{f}_2 } \biggr\}.
\label{eq:VVVC2}
\end{align}

\subsection{$\cA_{AAAH}$ amplitude}
\label{app:AAAH}
We can now embark for a long journey through the computation of $
\cA_{{A}{A}{A}H} $. 
With a judicious choice of super{-}ghost pictures and c{-}ghost insertions one has
\begin{align}
\cA(A^{({-}1)}_1,A^{(0)}_2,A^{(0)}_3,H^{({-}1)}_4)&  \nonumber\\
{=} \lim_{(z_1,z_2,z_4) \rightarrow (\infty, 1, 0)} \int_0^1 dz_3 \langle c e^{{-}\varphi} a_1\psi e^{ik_1X}(z_1) & c (a_2 \de X {+} i k_2\psi a_2\psi)e^{ik_2X}(z_2)\nonumber\\
&(a_3 \de X {+} i k_3\psi a_3\psi)e^{ik_3X}(z_3)
c \de X{\cdot}H_4{\cdot}\psi e^{ip_4X}(z_4)\rangle.
\end{align}
Since $\langle \psi(1) {:}\psi\psi\psi{:}(4)\rangle {=} 0$, there are only three kinds of contractions:
\begin{align}
&\langle \psi(1) \de X(3)  \de X(3) \psi\de X(4)\rangle,\\
&\langle \psi(1) {:}\psi\psi{:}(2)  \de X(3) \psi\de X(4)\rangle {+} (2\leftrightarrow 3),\\
&\langle \psi(1) {:}\psi\psi{:}(2)  {:}\psi\psi{:}(3) {:}\psi\de X(4)\rangle. 
\end{align}
Exploiting the $P_i(x)$ allows to identify 11 terms:
\begin{align}
&{-} a_1{\cdot}H_4{\cdot}a_2 a_3{\cdot}\left( {1\over x }k_1 {+} {1\over x(1{-}x) }k_2  \right) \rightarrow {-} a_1{\cdot}H_4{\cdot}a_2 [a_3{\cdot}k_1
\cB(1,1) {+} a_3{\cdot}k_2 \cB(1,0)] \\
&{1\over x^2} a_1{\cdot}H_4{\cdot}a_3 a_2{\cdot}\left( { x \over 1{-}x} k_3 {-} k_1\right)  \rightarrow
a_1{\cdot}H_4{\cdot}a_3 [a_2{\cdot}k_3 \cB(1,1) {-} a_2{\cdot}k_1 \cB(0,1) \\
 &{-} {1\over (1{-}x)^2} a_2{\cdot}a_3 a_1{\cdot}H_4{\cdot}\left(k_2 {+} { 1\over x} k_3 \right)\rightarrow  {-} a_2{\cdot}a_3 a_1{\cdot}H_4{\cdot}[k_2\cB(2,{-}1) {+} k_3\cB(1,{-}1)]\\
 &{1\over x} a_1{\cdot}f_3 {\cdot}H_4{\cdot}a_2 \rightarrow  a_1{\cdot}f_3 {\cdot}H_4{\cdot}a_2 \cB(1,1)\\
 &{1\over x^2} a_1{\cdot}f_2 {\cdot}H_4{\cdot}a_3 \rightarrow  a_1{\cdot}f_2 {\cdot}H_4{\cdot}a_3 \cB(0,1)
\end{align}
\begin{align}
& a_1{\cdot}H_4{\cdot}\left(k_2 {+} { 1 \over x} k_3\right) a_2{\cdot}\left( { x \over 1{-}x} k_3 {-} k_1\right)
a_3{\cdot}\left( {1\over x }k_1 {+} {1\over x(1{-}x) }k_2  \right)   \rightarrow\nonumber\\
&a_1{\cdot}H_4{\cdot}k_2\{a_2{\cdot}k_3 [a_3{\cdot}k_1 \cB(2,0) {+} a_3{\cdot}k_2 \cB(2,{-}1)]
{-} a_2{\cdot}k_1 [a_3{\cdot}k_1 \cB(1,1) {+} a_3{\cdot}k_2 \cB(1,0)]\} \nonumber\\
&{+} a_1{\cdot}H_4{\cdot}k_3
\{a_2{\cdot}k_3 [a_3{\cdot}k_1 \cB(1,0) {+} a_3{\cdot}k_2 \cB(1,{-}1)]
{-} a_2{\cdot}k_1 [a_3{\cdot}k_1 \cB(0,1) {+} a_3{\cdot}k_2 \cB(0,0)]\}\\
& {-}{1\over x} a_1{\cdot}f_3{\cdot}H_4{\cdot}\left(k_2 {+} { 1 \over x} k_3\right) a_2{\cdot}\left( { x \over 1{-}x} k_3 {-} k_1\right) \rightarrow   \nonumber\\
&{-} a_1{\cdot}f_3{\cdot}H_4{\cdot}k_2[a_2{\cdot}k_3 \cB(2,0) 
{-} a_2{\cdot}k_1 \cB(1,1)] {-} a_1{\cdot}f_3{\cdot}H_4{\cdot}k_3[a_2{\cdot}k_3 \cB(1,0) 
{-} a_2{\cdot}k_1 \cB(0,1)] \\
& a_1{\cdot}f_2{\cdot}H_4{\cdot}\left(k_2 {+} { 1 \over x} k_3\right) a_3\left( {1\over x }k_1 {+} {1\over x(1{-}x) }k_2  \right) \rightarrow 
\end{align}
\begin{align}
&a_1{\cdot}f_2{\cdot}H_4{\cdot}k_2[a_3{\cdot}k_1 \cB(1,1) 
{+}a_3{\cdot}k_2 \cB(1,0)] {+} a_1{\cdot}f_2{\cdot}H_4{\cdot}k_3[a_3{\cdot}k_1 \cB(0,1) 
{+}a_3{\cdot}k_2 \cB(0,0)]\\
& {1\over 1{-}x} a_1{\cdot}f_3{\cdot}f_2{\cdot}H_4{\cdot}\left(k_2 {+} { 1 \over x} k_3\right) \rightarrow   
a_1{\cdot}f_3{\cdot}f_2{\cdot}H_4{\cdot}k_2 \cB(2,0) {+}  
a_1{\cdot}f_3{\cdot}f_2{\cdot}H_4{\cdot}k_3 \cB(1,0)\\
&  {-} {1\over x(1{-}x)} a_1{\cdot}f_2{\cdot}f_3{\cdot}H_4{\cdot}\left(k_2 {+} { 1 \over x} k_3\right) \rightarrow   
{-} a_1{\cdot}f_2{\cdot}f_3{\cdot}H_4{\cdot}k_2 \cB(1,0) {-}  
a_1{\cdot}f_3{\cdot}f_2{\cdot}H_4{\cdot}k_3 \cB(0,0)
\end{align}
\begin{align}
&  {-} {1\over (1{-}x)^2} f_2{:}f_3 a_1{\cdot}H_4{\cdot}\left(k_2 {+} { 1 \over x} k_3\right) \rightarrow   
{-} f_2{:}f_3 [a_1{\cdot}H_4{\cdot}k_2 \cB(2,{-}1) {+}  
a_1{\cdot}H_4{\cdot}k_3 \cB(1,{-}1)
\end{align}
Factoring out 
$$
\cB(1,1) {=} {\Gamma(1{-}\ap s) \Gamma(1{-}\ap t) \over \Gamma(1 {+} \ap u) } 
$$ 
finally yields 
\begin{align}
&\cA_{{A} {A} {A} H} {=}\cB(1,1)\biggr\{ {-}\frac{u}{t}a_2a_3\biggr[ (1{-}\alp s)\,\,a_1H_4k_2 {+} (\alp u {-}1)\,\,a_1H_4k_3\biggr] {-}a_1H_4a_2\biggr(a_3k_1{-}a_3k_2\frac{u}{t}\biggr)\nonumber\\
&{-}a_1H_4a_3\biggr(a_2k_3\frac{u}{t}{-}a_2k_1\frac{u}{s}\biggr){+}2\alp a_1H_4k_2\biggr[{-}a_2k_3\,\,a_3k_1 \frac{1{-}\alp s}{\alp t}{-}a_2k_1\biggr(a_3k_1 {-}a_3k_2\frac{u}{t}\biggr)\biggr]\nonumber\\
&{+}2\alp a_1H_4k_3\biggr[{-}a_2k_3\,\,a_3k_1\frac{u}{t}{-}a_2k_1\biggr({-}a_3k_1\frac{u}{s}{+}a_3k_2\frac{u(\alp u {-}1)}{\alp s t}\biggr)\biggr]{+}a_1f_2H_4a_3\frac{u}{s}{-}a_1f_3H_4a_2\nonumber\\
&{-}2\alp a_1f_2H_4k_2\biggr(a_3k_1{-}a_3k_2\frac{u}{t}\biggr){-}2\alp a_1f_2H_4k_3\biggr({-}a_3k_1\frac{u}{s}{+}a_3k_2\frac{u(\alp u {-}1)}{\alp s t}\biggr)\nonumber\\
&{+}2\alp a_1f_3H_4k_2\biggr({-}a_2k_3\frac{1{-}\alp s}{\alp t}{-}a_2k_1\biggr) {+} 2\alp a_1f_3H_4k_3\biggr({-}a_2k_3\frac{u}{t}{+}a_2k_1\frac{u}{s}\biggr)\nonumber\\
&{+}2\alp a_1f_2f_3H_4k_2\frac{u}{t}{-}2\alp a_1f_2f_3H_4k_3\frac{u(\alp u {-}1)}{\alp s t}{-}2\alp a_1f_3f_2H_4k_2 \frac{1{-}\alp s}{\alp t}{-}2\alp a_1f_3f_2H_4k_3\frac{u}{t}\biggr\}.
\label{eq:AAAH0}
\end{align}
\subsection{$\cA_{AAAC}$ in 4{-}dimensions}
\label{AAAC_4D}
Let us first consider $\cA(1^{+}2^{+}3^{+}C)$.
 \begin{align}
&\Amp(1^{+}2^{+}3^{+}C){=}i \cB(1,1)\biggr\{a_1^{+}f_2^{+}p_4\biggr(a_3^{+} k_1 {-}\frac{u}{t}a_3^{+}k_2\biggr) {+} a_1^{+}f_3^{+}p_4\biggr(a_2^{+}k_3\frac{u}{t}{-}a_2^{+}k_1\frac{u}{s}\biggr)\nonumber\\
&{+}\frac{u}{s}a_1^{+}k_2\,a_2^{+}f_3^{+}p_4{+}a_1^{+}k_3\,a_3^{+}f_2^{+}p_4 {-} \biggr[ a_2^{+}k_3(a_3^{+}f_1^{+}p_4{-}a_1^{+}f_3^{+}p_4){-}a_3^{+}k_2(a_2^{+}f_1^{+}p_4{-}a_1^{+}f_2^{+}p_4)\biggr]\biggr\}.
 \end{align}
 The final result is 
\be \Amp(1^{+}2^{+}3^{+}C){=}i\cB(1,1)m_C^4\frac{[13]}{\lef 12\re\lef 23\re}.\ee

Let us now consider $\Amp(1^{-}2^{-}3^{+}C)$.
Choosing  \be
a_1^{{-}\mu}{=}\frac{ \lef 1| \sigma^\mu |2] }{[21]}\quad a_2^{{-}\mu}{=}\frac{ \lef 2| \sigma^\mu |1] }{[12]}\quad  a_3^{{+}\mu}{=}\frac{ \lef 1| \sigma^\mu |3] }{\lef 13 \re},
\ee
the amplitude simplifies as follows
\begin{align}
\Amp(1^{-}2^{-}3^{+}C)&{=}i\biggr({-}a_1^-{{\cdot}} C{:}f^+_3\,\,a_2^-k_3\frac{u}{t}{-}a_1^-f_3^+{\cdot} C{:}f^-_2{-}2\frac{u}{t}a^-_1{\cdot} \dot{C}\overline{{\cdot} f^+_3{\cdot} }\dot{f}_2^-\biggr)\cB(1,1)\\
&{=}ia_1^-k_3\,\,a_2^-k_3\,\,a_3^+k_2\,\cB(1,1){=}i\frac{[13]}{\lef 12\re } \lef 23\re^3\cB(1,1).
\end{align}
\subsection{$\cA_{{A}{A}{A}H}$ in 4{-}dimensions}
\label{AAAH_D4}
Expressing the 4{-}momentum of a $H$ massive spin 2 state as $p{=}k_1{+}k_2$, with $k_1^2{=}k_2^2{=}0$ and $2k_1{\cdot} k_2{=}p^2{=}{-}1/\alp$, it is possible to 
write its physical polarizations in the spinor helicity formalism. If we define $k_{1\al\ald}{=}u_\al \bar u_{\ald}$ and $k_{2\al\ald}{=}v_\al \bar v_{\ald}$, 
we have
\begin{align}
H_{\al\ald\beta\betad}&{=}c_0(u_\al u_\beta \bar u_{\ald} \bar u_{\betad} {+} v_\al v_\beta \bar v_{\ald } \bar v_{\betad} {-} (u_\al v_\beta {+} u_\beta v_\al)(\bar u_{\ald} \bar v_{\betad} {+} \bar u_{\ald} \bar v_{\betad}))\nonumber\\
&{+}c_1(u_\al u_\beta (\bar u_{\ald} \bar v_{\betad} {+} \bar u_{\betad} \bar v_{\ald} ) {-} \bar v_{\ald} \bar v_{\betad} (u_\al v_\beta {+} u_\beta v_\al) )\nonumber\\
&{+}c_{{-}1}(v_\al v_\beta (\bar u_{\ald} \bar v_{\betad} {+} \bar u_{\betad} \bar v_{\ald} ) {-} \bar u_{\ald} \bar u_{\betad} (u_\al v_\beta {+} u_\beta v_\al) )\nonumber\\
&{+}c_2 u_\al u_\beta \bar v_{\ald} \bar v_{\betad} {+} c_{{-}2} v_\al v_\beta \bar u_{\ald} \bar u_{\ald}.
\end{align}
Recalling that $\cA(A_1,A_2,H)\propto \tr(f_1Hf_2)$, we can express the coupling between two vector bosons and each helicity component of $H$.
As shown in Tab.~\ref{tab:VVH}, $H$ couples only to vector bosons with opposite helicity.

\begin{table}
\begin{center}
\begin{tabular}{|c||c|c|c|c|c|}
\hline
 & $f_1^{-}\,f_2^{+}$ & $f_1^{-}\,f_2^{-}$ & $f_1^{+} \,f_2^{+}$ & $f_1^{+} \,f_2^{-}$ \\
\hline\hline
$c_0$ & $\lef u1 \re^2[u2]^2 {+} \lef v1 \re^2 [v2]^2$ & 0 & 0 &  $\lef u2 \re^2[u1]^2 {+} \lef v2 \re^2 [v1]^2$\\
 & ${-}4\lef u1 \re \lef v1 \re [u2][v2] $& & & ${-}4\lef u2 \re \lef v2 \re [u1][v1]$\\\hline
$c_1$ & $2 \lef u1 \re^2 [u2][v2] {-}2 \lef u1 \re \lef v1 \re [v2]^2$ & 0 & 0 & $2 \lef u2 \re^2 [u1][v1] {-}2 \lef u2 \re \lef v1 \re [v1]^2$\\\hline
$c_{{-}1}$ & $2 \lef v1 \re^2 [v2][u2] {-}2 \lef v1 \re \lef u1 \re [u2]^2$ & 0 & 0 & $2 \lef v2 \re^2 [v1][u1] {-}2 \lef v2 \re \lef u1 \re [u1]^2$\\\hline
$c_2$ & $\lef u1 \re^2 [v2]^2$ & 0 & 0 & $\lef u2 \re^2 [v1]^2 $\\\hline
$c_{{-}2}$ & $\lef v1 \re^2 [u2]^2 $& 0 & 0 & $\lef v2 \re^2 [u1]^2$\\
\hline
\end{tabular}
\end{center}
\caption{In the table we list all couplings between a spin-2 massive state $H$ and two vector bosons $A_1$, $A_2$. The momentum of $H$ is $p_{\alpha\dot\alpha}{=}u_{\alpha}\bar u_{\dot\alpha}{+}v_{\alpha}\bar v_{\dot\alpha}$. It is worth to notice that $H$ couples only to couple of vector bosons with opposite helicities.}
\label{tab:VVH}
\end{table}
Let us discuss the case in which only the scalar component of $H$ (the trace of $H$ in 4 dimensions) couples to the three vector bosons and 
let us start with the amplitude $\cA(1^{+}2^{+}3^{+}H_0)$.
Choosing 
\be
a_1^{{+}}{=}\frac{  |1]\lef 2|  }{\lef 21 \re}\quad a_2^{{+}}{=}\frac{  |2]\lef 1|  }{\lef 12 \re}\quad  a_3^{{+}}{=}\frac{  |3]\lef 1| }{\lef 13 \re},
\ee
some of the scalar products appearing in Eq.~\eqref{eq:AAAH0} vanish: $a_2^{+}{\cdot} a_3^{+}{=}0$, $a_1^{+}{\cdot} k_2{=}0$, $a^{+}_2{\cdot} k_1{=}0$ and $a^{+}_3{\cdot} k_1{=}0$, and 
the amplitude simplifies significantly
\begin{align}
&\cA(1^{+}2^{+}3^{+}H_0){=}a_1H_0a_2\,\,a_3k_2\frac{u}{t}{-}a_1H_0a_3\,\,a_2k_3\frac{u}{t}{+}a_1f_2H_0a_3\frac{u}{s}{-}a_1f_3H_0a_2{+}2\alp a_1f_2H_0k_2\,\,a_3k_2\frac{u}{t}\nonumber\\
&{+}2\alp a_1f_2H_0k_3\,\,a_3k_2\frac{u(s{+}t)}{st}{-}2\alp a_1f_3H_0k_2\,\,a_2k_3\frac{u{+}t}{t}{-}2\alp a_1f_3H_0k_3\,\,a_2k_3\frac{u}{t}\nonumber\\
&{+}2\alp a_1f_2f_3H_0k_2\frac{u}{t}{+}2\alp a_1f_2f_3H_0k_3\frac{u(s{+}t)}{st}{-}2\alp a_1f_3f_2H_0k_2\frac{u{+}t}{t}{-}2\alp a_1f_3f_2H_0k_3\frac{u}{t}.
\label{VVVH0}
\end{align}
The scalar 4-dimensional polarization of $H_0$ is $H_{0\mu\nu}{=}\eta_{\mu\nu}{+}\alp p_\mu p_\nu$.
Let us consider the diagonal part of the polarization of $H_0$.
\begin{align}
a_1a_2\,\,a_3k_2\frac{u}{t}{-}a_1a_3\,\,a_2k_3\frac{u}{t}&{=}2(s{+}t)\frac{[13]}{\lef 12 \re \lef 23 \re }\nonumber\\
a_1f_2a_3\frac{u}{s}{-}a_1f_3a_2&{=}4t\frac{[13]}{\lef 12 \re \lef 23 \re}\nonumber\\
2\alp a_1f_2k_3\,\,a_3k_2\frac{u(s{+}t)}{st}&{=}{-}2\alp t (s{+}t)\frac{[13]}{\lef 12 \re \lef 23 \re}\nonumber\\
{-}2\alp a_1f_3f_2k_3\frac{u}{t}&{=}{-}2\alp ut\frac{[13]}{\lef 12 \re \lef 23 \re}
\label{1H0}
\end{align}
Let us now consider the longitudinal part of the polarization: $\alp p_\mu p_\nu$.
\begin{subequations}\begin{align}
\alp a_1p\,\,a_2p\,\,a_3k_2\frac{u}{t}{-}\alp a_1p\,\,a_3p\,\,a_2k_3\frac{u}{t}&{=}0\\
\alp a_1f_2p\,\,a_3p\frac{u}{s}{-}\alp a_1f_3p\,\,a_2p&{=}{-}t\frac{[13]}{\lef 12 \re \lef 23 \re}\\
2\alpha^{\prime 2} a_1f_2p\,\,pk_2\,\,a_3k_2\frac{u}{t}{+}2\alpha^{\prime 2} a_1f_2p\,\,pk_3\,\,a_3k_2\frac{u(s{+}t)}{st}&{=}\alp(s{+}t)^2\frac{[13]}{\lef 12 \re \lef 23 \re}\\
{-}2\alpha^{\prime2} a_1f_3p\,\,pk_2\,\,a_2k_3\frac{u{+}t}{t}{-}2\alpha^{\prime 2} a_1f_3p\,\,pk_3\,\,a_2k_3\frac{u}{t}&{=}\alp t(u{+}t)\frac{[13]}{\lef 12 \re \lef 23 \re}\\
2\alpha^{\prime 2} a_1f_2f_3p\,\,pk_2\,\,\frac{u}{t}{+}2\alpha^{\prime 2} a_1f_2f_3p\,\,pk_3\,\,\frac{u(s{+}t)}{st}&{=}\alp u(s{+}t)\frac{[13]}{\lef 12 \re \lef 23 \re}\\
{-}2\alpha^{\prime 2} a_1f_3f_2p\,\,pk_2\frac{u{+}t}{t}{-}2\alpha^{\prime 2}a_1f_3f_2p\,\,pk_3\frac{u}{t}&{=}\alp(s{+}t)(u{+}t)\frac{[13]}{\lef 12 \re \lef 23 \re}.
\end{align}\label{2H0}\end{subequations}
Using the identity $s{+}t{+}u{=}1/\alp$, the sum of the terms in Eqs.~\eqref{1H0},~\eqref{2H0} yields
\be\cA(1^{+}2^{+}3^{+}H_0)\propto  \frac{1}{\alp}\frac{[13]}{\lef 12 \re \lef 23\re}. \ee

Let us consider the amplitude $\Amp(1^{-}2^{-}3^{+}H_0)$.
Choosing \be
a_1^{{-}}{=}\frac{  |2] \lef 1|}{[21]}\quad a_2^{{-}}{=}\frac{  |1]\lef 2|  }{[12]}\quad  a_3^{{+}}{=}\frac{  |3]\lef 1| }{\lef 13 \re},
\ee
we can enforce the conditions $a_i{\cdot} a_j{=}0$ and $a_1{\cdot} k_2{=}a_2{\cdot} k_1{=}a_3{\cdot} k_1{=}0$.
The resulting amplitude looks like
\begin{align}
&\Amp(1^{-}2^{-}3^{+}H_0){=}a_1H_0a_2\,\,a_3k_2\frac{u}{t}{+}a_1H_0a_3\,\,a_2k_3\frac{u}{t}{+}a_1k_3\,\,a_3H_0a_2{+}2\alp a_1k_3\,\,a_3H_0k_2\,\,a_2k_3\frac{u{+}t}{t}\nonumber\\
&{+}2\alp a_1k_3\,\,a_3H_0k_3\,\,a_2k_3\frac{u}{t}{-}2\alp a_1k_3\,\,a_3k_2\,\,a_2H_0k_2\frac{u{+}t}{t}{-}2\alp a_1k_3\,\,a_3k_2\,\,a_2H_0k_3\frac{u}{t}.
\end{align}
The diagonal part of the polarization of $H_0$ produces
\begin{align}
2\alp a_1k_3\,\,a_3H_0k_2\,\,a_2k_3\frac{u{+}t}{t}{-}2\alp a_1k_3\,\,a_3k_2\,\,a_2H_0k_3\frac{u}{t}{=}2\alp \frac{[13]\lef 23\re^3 }{\lef 12 \re}.
\end{align}
The longitudinal part of the polarization of $H_0$ yields
\begin{align}
&\alp a_1p\,\,a_2p\,\,a_3k_2\frac{u}{t}{+}\alp a_1p\,\,a_3p\,\,a_2k_3\frac{u}{t}{+}\alp a_1k_3\,\,a_3p\,\,a_2p{=}3\alp \frac{[13]\lef 23\re^3}{\lef 12\re}
\end{align}
Finally the result is 
\be \Amp(1^{-}2^{-}3^{+}H_0)\propto \alp\frac{[13]}{\lef 12\re}\lef 23\re^3. \ee

Let us consider the case in which the spin{-}2 tensor $H$ with helicity $u_\alpha(4)u_\alpha(4)\bar v_{\dot\alpha} (5) \bar v_{\dot\alpha}(5)$, with $p{=}k_4{+}k_5$ the momentum of $H$, having helicity $h{=}2$ couples to three vector bosons, \emph{i.e.} 
$\cA(1^{-}2^{+}3^{+}H)$.
Choosing the following parametrization for the polarization vectors of the incoming gluons
\be
a_1^{-}{=}\frac{ | 2]\lef 1 |}{[21]},\quad a_2^{+}{=}\frac{ | 2]\lef 1 | }{\lef 12 \re},\quad a_3^{+}{=}\frac{| 3]\lef 1 | }{\lef 13 \re},
\ee
we have $a_i{\cdot} a_j{=}0$ and $a_1^{-}{\cdot} k_2{=}a_2^{+}{\cdot} k_1{=}a_3^{+}{\cdot} k_1{=}0$. 
\begin{align}
&\cA(1^{-}2^{+}3^{+}H){=}\nonumber\\
&a_1^{-}Ha_2^{+}\,\,a_3^{+}k_2\frac{u}{t}{-}a_1^{-}Ha_3^{+}\,\,a_2^{+}k_3\frac{u}{t}{-}a_1^{-}f_3^{+}Ha_2^{+}{-}2\alp a_1^{-}f_3^{+}Hk_2\,\,a_2^{+}k_3\frac{u{+}t}{t}\nonumber\\
&{-}2\alp a_1^{-}f_3^{+}Hk_3\,\,a_2^{+}k_3\frac{u}{t}{-}2\alp a_1^{-}f_3^{+}f_2^{+}Hk_2\frac{u{+}t}{t}{-}2\alp a_1^{-}f_3^{+}f_2^{+}Hk_3\frac{u}{t}\nonumber\\
&{=}a_1^{-}Ha_2^{+}\,\,a_3^{+}k_2\frac{u}{t}{-}a_1^{-}Ha_3^{+}\,\,a_2^{+}k_3\frac{u}{t} {+} a_1^{-}k_3\,\,a_3^{+}Ha_2^{+} {+} 2\alp a_1^{-}k_3\,\, a_3^{+} Hk_2\,\,a_2^{+}k_3\frac{u{+}t}{t}\nonumber\\
&{+}2\alp a_1^{-} k_3\,\,a_3^{+} Hk_3\,\,a_2^{+}k_3\frac{u}{t}{-}2\alp a_1^{-}k_3\,\,a_3^{+}k_2\,\, a_2^{+} Hk_2\frac{u{+}t}{t}{-} 2\alp a_1^{-} k_3\,\,a_3^{+}k_2\,\,a_2^{+} Hk_3\frac{u}{t}.
\end{align}
So we have
\begin{align}
&a_1^{-}Ha_2^{+}\,\,a_3^{+}k_2\frac{u}{t}{-}a_1^{-}Ha_3^{+}\,\,a_2^{+}k_3\frac{u}{t}{=}4\cA\frac{[25][45]\lef 45 \re^2}{[21]\lef 14 \re}\nonumber\\
&a_1^{-}k_3\,\,a_3^{+}Ha_2^{+}{=}4\cA\frac{[23][25][35]\lef 45 \re^2\lef 23 \re}{[21][13]\lef 14 \re^2}\nonumber\\
&2\alp a_1^{-}k_3\,\, a_3^{+} Hk_2\,\,a_2^{+}k_3\frac{u{+}t}{t}{+}2\alp a_1^{-} k_3\,\,a_3^{+} Hk_3\,\,a_2^{+}k_3\frac{u}{t}\nonumber\\
&{=}{-}4\alp\cA\frac{[23][35]\lef 13 \re \lef 45 \re^2}{[21][13]\lef 14 \re^3}(t[25]\lef 24 \re {-} u[15]\lef 14\re)\nonumber\\
&{-}2\alp a_1^{-}k_3\,\,a_3^{+}k_2\,\, a_2^{+} Hk_2\frac{u{+}t}{t}{-} 2\alp a_1^{-} k_3\,\,a_3^{+}k_2\,\,a_2^{+} Hk_3\frac{u}{t}\nonumber\\
&{=}{-}4\alp\cA\frac{[23][25]\lef 12 \re \lef 45 \re^2}{[21][13]\lef 14 \re^3}(t[25]\lef 24 \re {-} u[15]\lef 14\re),
\label{HD}
\end{align}
where
\be \cA{=}\frac{\lef 14\re^4 [13]}{\lef 12 \re \lef 23 \re \lef 45 \re^2}. \ee
The sum of the terms in Eq.~\eqref{HD}, produces the amplitude
\be \cA(1^{-}2^{+}3^{+}H)\propto\frac{1}{\alp}\cA{=}\frac{1}{\alp}\frac{\lef 14 \re^4 [13]}{\lef 12 \re \lef 23 \re \lef 45 \re^2}.\ee

\end{document}